%% file: Bethlem_AG.tex
\documentclass[twocolumn,showpacs,preprintnumbers,amsmath,amssymb]{revtex4}

\usepackage{graphicx}
\usepackage{dcolumn}
\usepackage{bm}
\usepackage{threeparttable}
\usepackage{epsfig}

\begin{document}

\preprint{}

\title{Alternating Gradient Focusing and Deceleration of Polar Molecules}

\author{Hendrick L. Bethlem$^{1,2}$}
%%\altaffiliation[]{}
\email{rick@fhi-berlin.mpg.de}
\author{M.R. Tarbutt$^{3}$}
\author{Jochen K\"{u}pper$^{1}$}
\author{David Carty$^{1}$\footnote{Present
adress: Physical and Theoretical Chemistry Laboratory, South Parks Road,
Oxford OX1 3QZ, United Kingdom}}
\author{Kirstin Wohlfart$^{1}$}
\author{E.A. Hinds$^{3}$}
\author{Gerard Meijer$^{1}$}

\affiliation{
$^{1}$Fritz-Haber-Institut der Max-Planck-Gesellschaft,
Faradayweg 4-6, D-14195 Berlin, Germany\\
$^{2}$Laser Centre Vrije Universiteit,
De Boelelaan 1081, NL-1081HV Amsterdam, The Netherlands\\
$^{3}$Centre for Cold Matter, Blackett Laboratory, Imperial
College London, SW7 2BW, United Kingdom\\
}

\date{\today}

\begin{abstract}
Beams of polar molecules can be focused using an array of electrostatic lenses
in alternating gradient (AG) configuration. They can also be accelerated or
decelerated by applying an appropriate high voltage switching sequence to the
lenses. AG focusing is applicable to molecules in both low-field and
high-field-seeking states and is particularly well suited to the problem of
decelerating heavy molecules and those in their ground rotational state. We
describe the principles of AG deceleration and set out criteria to be followed
in decelerator design, construction and operation. We calculate the
longitudinal and transverse focusing properties of a decelerator, and exemplify
this by 2D-imaging studies of a decelerated beam of metastable CO molecules.
\end{abstract}


\maketitle

\section{Introduction}

During the last few years, a variety of techniques have been demonstrated to
produce samples of trapped neutral molecules \cite{Doyle:EPJD31:149:2004}. One
of these exploits the force that a polar molecule experiences in an
inhomogeneous electric field to change its motion. This force attracts
molecules to regions of high or low electric field depending on the sign of the
Stark shift. Some small polar molecules in low-field seeking states have been
decelerated using a series of pulsed electric fields. These include CO
\cite{Bethlem:PRL83:1558:1999}, NH$_{3}$ and ND$_{3}$
\cite{Bethlem:PRA65:053416:2002}, OH \cite{Bochinski:PRL91:243001:2003,
vandeMeerakker:PRL94:023004:2004}, NH \cite{vandeMeerakker:arXiv:2005},
H$_{2}$CO \cite{eHudson:arXiv:2005} and SO$_{2}$ \cite{Lisdat:Submitted}. In
the case of ND$_{3}$ \cite{Bethlem:PRA65:053416:2002} and OH
\cite{vandeMeerakker:PRL94:023004:2004} packets of Stark decelerated molecules
have subsequently been electrostatically trapped. We aim to extend this
deceleration method to heavy polar molecules including bio-molecules. Of
particular interest are molecules such as YbF which are being used in
experiments aimed at detecting time-reversal symmetry violating interactions
leading to a permanent electric dipole moment (EDM) of the electron, which is a
sensitive probe for physics beyond the Standard Model
\cite{Hudson:PRL89:023003:2002, jHudson:laserspec17:129:2005}. Decelerated
molecules offer an increased sensitivity for these experiments. Stark
deceleration of bio-molecules allows one to prepare samples of selected
conformers for further studies.

Deceleration of heavy polar molecules is difficult for two reasons: (i) For a
given velocity of the beam, the kinetic energy of the molecules, and thus the
number of electric field stages required to bring the molecules to a
standstill, is proportional to their mass. (ii) At the electric field
strengths  required for deceleration, all low-lying rotational levels of heavy
molecules have a negative Stark shift. In these states the molecules are
attracted to a maximum of the electric field, i.e. to the electrodes. In order
to guide high-field-seeking molecules through the decelerator dynamic focusing
schemes need to be used which, typically, have an order of magnitude smaller
acceptance than the schemes used to guide low-field-seeking molecules.  We have
recently demonstrated a decelerator for high-field seeking molecules
\cite{Bethlem:PRL88:133003:2002, Tarbutt:PRL92:173002:2004} using the
alternating gradient principle. Those experiments showed the feasibility of the
AG deceleration technique, which we discuss in detail here.

Our paper is organized as follows. In Sec.~\ref{Sec:GeneralPrinciples} we
outline the method of alternating gradient Stark deceleration, and set out some
general principles that will guide us through the rest of the paper. In
Sec.~\ref{Sec:Stark} the Stark shift of polar molecules is discussed in more
detail, drawing on the particular cases of metastable CO, YbF and benzonitrile.
In Sec.~\ref{Sec:geometry} we present three simple electrode geometries that
may be used to make a single lens of the alternating gradient array and discuss
the merits of these geometries. In Sec.~\ref{Sec:AG} we consider the motion of
the molecules through the decelerator, present the trajectories and phase-space
distributions, and calculate the transmission of both idealised and real
decelerators. In Sec.~\ref{Sec:experimental} we present an experimental study
of the transverse focusing properties of an AG decelerator, by measuring the
two-dimensional distributions of a decelerated beam of metastable CO molecules.
Our results are compared to calculations. A summary of our main conclusions and
a discussion of future prospects are given in Sec.~\ref{Sec:conclusion}.

\section{\label{Sec:GeneralPrinciples}
General Principles}

\setlength{\epsfxsize}{0.45\textwidth}
\begin{figure}
\centerline{\epsffile{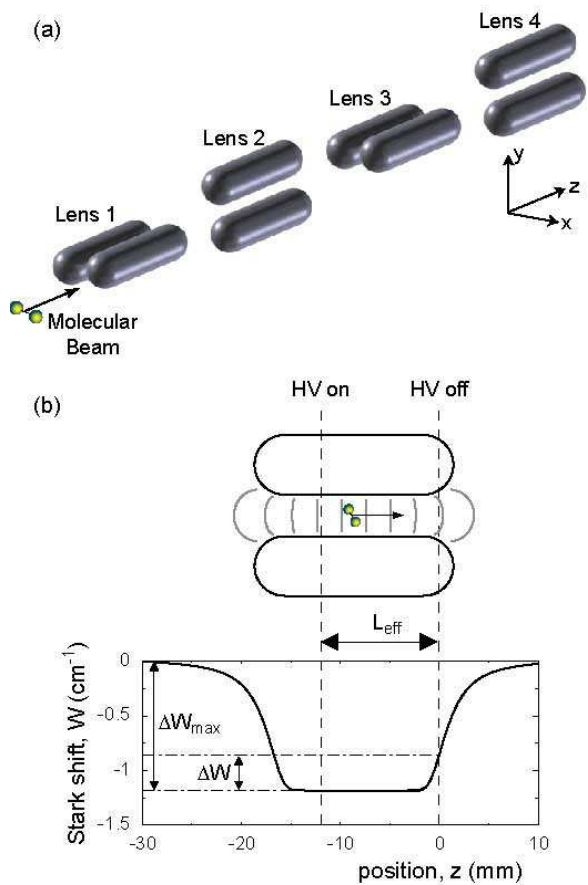}}
\protect
\vspace*{0.5cm}
\caption{\label{Fig:ExperimentLayout} (a) Layout of an alternating gradient
decelerator for polar molecules showing the first four deceleration stages.
Each electrode pair acts both to focus and decelerate the molecules. (b)
Cross-section of a single lens formed from two 20\,mm long rods with
hemispherical ends, 6\,mm in diameter and spaced 2\,mm apart. Potential energy
along the $z$-axis for metastable CO molecules in the a$^{3}\Pi$, $J$=1,
$\Omega$=1, $M\Omega$=+1 level, when the potential difference between the
electrodes is 20\,kV. The high-voltage switching procedure is indicated: the
voltages are turned on when the bunch of molecules reaches the `HV on'
position, and are turned off once they reach the `HV off' position.}
\end{figure}

A Stark decelerator consists of a series of capacitor plates. A
polar molecule that has its dipole oriented anti-parallel to the
electric field (a low-field seeker) will gain potential energy
when entering one of these capacitors and will therefore be
decelerated. When leaving the field of the capacitor it will lose
potential energy and so be accelerated back to its initial
velocity. If we switch the field off before the molecule has left
the capacitor, it will keep its lower velocity. Similarly, a
molecule that has its dipole oriented parallel to the electric
field (a high-field seeker) will accelerate when entering, and
decelerate when leaving one of these capacitors. Again, by
switching the electric fields at the appropriate times the
molecule will be decelerated. A series of switched electric fields
can thus be used to decelerate a pulsed molecular beam. The
initial velocity of a seeded supersonic beam of molecules is in
the range 250--2000\,m/s, depending on the mass and temperature of
the carrier gas. Typically, more than 100 stages are required to
decelerate these beams to zero velocity. In order to have useful
transmission, it is therefore of utmost importance that the
trajectories through the decelerator are stable.

For a force field, $\vec{F}(\vec{r})$, to keep a particle in static equilibrium
around $\vec{r}=0$, two conditions must be met. The applied force must vanish
at $\vec{r}=0$, and, for small displacements, the force field should tend to
restore the particle towards $\vec{r}=0$ \footnote{We consider focusing in
three directions here. It is possible to create fields that focus molecules in
high-field-seeking states in the transverse direction while defocusing them in
the longitudinal direction. These fields cannot be used for phase-stable
deceleration.}. To achieve the latter it is necessary that the divergence of
the force be negative, $\vec{\nabla}\cdot\vec{F}<0$ \footnote{The condition
$\vec{\nabla}\cdot\vec{F}<0$, is necessary but not sufficient for
stability. For instance, the force may vary in such a way that parametric
amplification of the amplitude of the trajectories occurs (see, for instance,
\cite{vandeMeerakker:PRA73:023401:2006}).}. The force acting on the molecules
in an inhomogeneous electric field is given by:

\begin{equation}
\vec{F}\left(\vec{r}\right) = -\vec{\nabla}  W(E),
\label{Eq:dipoleforce}
\end{equation}

\noindent with $W(E)$ being the Stark shift of a polar molecule in
an electric field of magnitude $E=|\vec{E}|$. The properties of
this force field were analyzed in a seminal paper by Auerbach,
Bromberg and Wharton \cite{Auerbach:JCP45:2160:1966}. For
molecules that experience a linear Stark shift in the applied
field, $W= -\mu_{\mathrm{\it{eff}}} E$, it was shown that:

\begin{equation}
\begin{split}
\vec{\nabla}\cdot \vec{F} = \frac{\mu_{\mathrm{\it{eff}}}}{E^{3}}
&\sum_{i,j,k=1}^{3} \left[\left(\frac{\partial\Phi}{\partial
x_{k}}\right)^{2} \left(\frac{\partial^{2}\Phi}{\partial x_{i}
\partial x_{j}}\right)^{2}  \right.\\
 &- \left. \left(\frac{\partial \Phi}{\partial x_{i}}\right)
\left(\frac{\partial \Phi}{\partial x_{k}}\right)
\left(\frac{\partial^{2} \Phi}{\partial x_{i} \partial
x_{j}}\right) \left(\frac{\partial^{2} \Phi}{\partial x_{k}
\partial x_{j}}\right) \right],
\end{split}
\label{Eq:divergenceOfForce}
\end{equation}

\noindent
where $\Phi$ is the electrostatic potential and $\mu_{\mathrm{\it{eff}}}$ is an
effective dipole moment which depends on the particular molecular state. Using
Schwartz's inequality, it can be seen that the sum is always positive.
Therefore, for molecules having a linear Stark shift the sign of
$\vec{\nabla}\cdot\vec{F}$ is determined solely by the sign of
$\mu_{\mathrm{\it{eff}}}$ \footnote{A similar expression can be derived for a
quadratic Stark shift \cite{Auerbach:JCP45:2160:1966}.}. Thus, for molecules
that have a negative $\mu_{\mathrm{\it{eff}}}$ (low-field seekers),
$\vec{\nabla}\cdot\vec{F} \leq 0$, and focusing is straightforward. For
molecules that have a positive $\mu_{\mathrm{\it{eff}}}$ (high-field seekers),
$\vec{\nabla}\cdot\vec{F} \geq 0$, and focusing is more problematic.

The difficulty of focusing high-field-seeking molecules is analogous to the
situation for ions, for which
$\vec{\nabla}\cdot\vec{F}=q\vec{\nabla}\cdot\vec{E}=0$ in free space, where $q$
is the charge of the ion.  Therefore, techniques routinely applied to ions can
be translated to polar molecules. Three schemes are generally employed. (i)
{\it Circular motion}; In a cyclotron the curvature of the trajectory adds a
force which, in an appropriately shaped magnetic or electric field, stabilizes
the motion of the ions \cite{Livingood:Book:1961}. A similar stabilization can
be achieved for polar molecules. For example, molecules that have a linear
Stark shift in an applied field flying at a distance $r$ from the axis of a
capacitor formed by two coaxial cylinders, experience a force proportional to
1/$r^{2}$. They therefore move in stable Kepler-type orbits around the central
electrode. This technique has been used to focus molecules in high-field
seeking states
\cite{Helmer:JAP31:458:1960,Chien:CP7:161:1974,Loesch:PRL85:2709:2000}.  (ii)
{\it Alternating gradient (AG) focusing}; Alternating gradient focusing of
charged particles was pioneered by Courant, Livingstone and Snyder
\cite{Courant:PR88:1190:1952,Courant:AnnPhys3:1:1958} and is now applied in
virtually all particle accelerators. An AG array consists of a series of
magnetic or electric quadrupole lenses that focus ions in one direction while
defocusing them in the other direction. By alternating the orientation of these
fields it is possible to obtain net focusing in both directions. As this
stabilization is due to the motion of the ion itself, it is referred to as
`dynamic' stability. Application of the technique to focus polar molecules was
demonstrated experimentally by Kakati and Lain\'e~\cite{Kakati}, by G\"{u}nther
et al.~\cite{Gunther,Lubbert1,Lubbert2} and by Bromberg
\cite{Bromberg:thesis:1972}. More recently, the AG technique was used to focus
metastable argon atoms released from a magneto-optical
trap~\cite{Noh:PRA61:041601(R):2000} and cesium atoms in an atomic
fountain~\cite{Kalnins:PRA72:043406:2005}. Furthermore, the transmission of
methylfluoride molecules through a 15\,m long AG beamline was modelled and
optimized~\cite{Kalnins:RSI73:2557:2002}. (iii) {\it Einzel lens}; In an Einzel
or uni-potential lens an ion is subjected to an acceleration along the axial
direction followed by an equal deceleration. In the radial direction ions are
focused on entering the fringe field and defocused on leaving it. This results
in a net (dynamic) focusing effect. The focusing is only effective when the
change in kinetic energy of the ions is a substantial fraction of their initial
energy and so Einzel lenses are only useful for low energy ion beams. A similar
effect is obtained for polar molecules entering and leaving a field region, but
again its usefulness is restricted to very low energy beams, e.g. for loading
slow molecules into a trap.

For a decelerator (or accelerator) for polar molecules alternating gradient
focusing seems the obvious choice. Fig.~\ref{Fig:ExperimentLayout}(a) shows the
general form of the experimental setup. The AG lenses are formed from a pair of
cylindrical electrodes to which a voltage difference is applied. Molecules will
be defocused in the plane containing the electrodes while being focused in the
orthogonal plane. As the molecules move down the beamline the focusing and
defocusing directions alternate. The defocusing lenses have a smaller effect on
the molecules than the focusing lenses, not because their power is smaller (it
is not), but because the molecules tend to be close to the axis when they are
inside the defocusing lenses and further away from the axis when they encounter
the focusing lenses. Molecules in high-field seeking states are accelerated
while entering the field of an AG lens and are decelerated while leaving the
field. By simply switching the lenses on and off at the appropriate times, AG
focusing and deceleration of polar molecules can be achieved simultaneously.
Fig.~\ref{Fig:ExperimentLayout}(b) shows the potential energy (the Stark shift)
along the $z$-axis of a single lens for a representative high-field seeking
molecule. The molecules enter each lens with the electric fields turned off so
that their speed is unchanged as they enter. The fields are then suddenly
turned on, and the high-field seeking molecules are decelerated as they leave
the lens and move from a region of high field to one of low field. This process
is repeated until the molecules reach the desired speed. As indicated in the
figure, the amount of deceleration can be controlled by choosing how far up the
potential hill the bunch of molecules has climbed before the fields are turned
off (i.e. by moving the `HV off' point in Fig.~\ref{Fig:ExperimentLayout}(b)).
Similarly, the effective length of each lens, $L_{\mathrm{\it{eff}}}$, can be
controlled by varying the amount of time that the fields are on (i.e. by moving
the `HV on' point in Fig.~\ref{Fig:ExperimentLayout}(b)).

A prototype machine of this type has been used to decelerate high-field seeking
metastable CO molecules from 275\,m/s to 260\,m/s
\cite{Bethlem:PRL88:133003:2002}. More recently, an improved device decelerated
ground state YbF molecules from 287\,m/s to 277\,m/s, corresponding to a 7\%
reduction of the kinetic energy \cite{Tarbutt:PRL92:173002:2004}.  Since then,
YbF and CaF molecules have been decelerated using an array of 21 lenses at
Imperial College London, while at the Fritz-Haber-Institut in Berlin, CO and
benzonitrile molecules have been decelerated using an array of 27 lenses. These
results will be presented elsewhere.

\section{\label{Sec:Stark}
The Stark shift in polar molecules}

\setlength{\epsfxsize}{0.35\textwidth}
\begin{figure}
\centerline{\epsffile{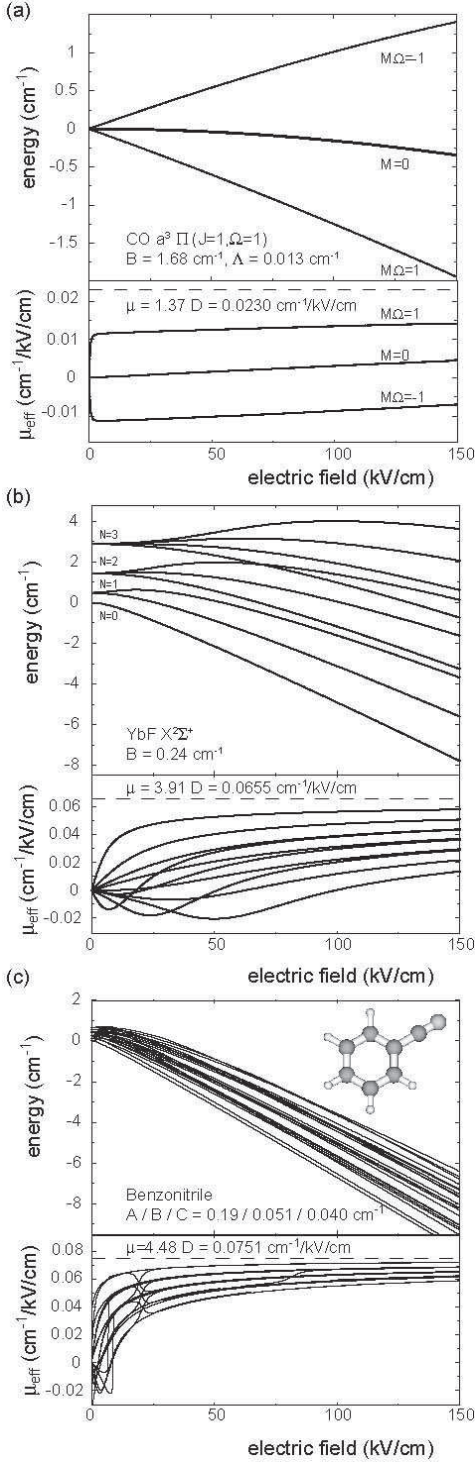}}
\protect
\vspace*{0.5cm}
\caption{\label{Fig:Stark} Stark shifts and effective dipole
moments for (a) the $J$=1, $\Omega$= 1 levels of the a$^{3}\Pi$
state of CO, (b) the lowest four rotational levels in the ground
state of YbF and (c) the lowest eight rotational levels of
benzonitrile. In (a) and (b) the basis set includes all levels up
to $J$=10, while in (c) all levels up to $J$=30 were included.}
\end{figure}

We now discuss the Stark shift in more detail for some representative polar
molecules. The Stark shift of a molecule is a function of the electric field
magnitude $E=|\vec{E}|$. It is useful to define a dimensionless parameter,
$\lambda$, that describes the strength of the electric field. In the context of
an idealized rigid-rotor molecule~\cite{Peter:JCP26:1657:1957} with dipole
moment $\mu$ and rotational constant $B$ (in energy units), the appropriate
dimensionless ratio is $\lambda=\mu E /B$. When the electric field is `weak',
$\lambda \ll 1$, the Stark shift is quadratic in $\lambda$ and can
be calculated using second order perturbation theory. The states are best
labelled by the rotational angular momentum quantum number $J$, and its
projection, $M$, onto the field axis. If $\lambda$ is increased, states of
different $J$ are increasingly strongly mixed until, in the strong-field limit
($\lambda \gg 1$), the states are called `pendular'
\cite{Loesch:JCP93:4779:1990, Friedrich:ZPhysD18:153:1991,
Rost:PRL68:1299:1992}. In that case they are labelled by the quantum numbers
$v_{p}$ and $M$, with $v_{p}=2J-|M|$, and states of the same $v_{p}$ but
different $M$ are degenerate. In this strong-field limit, all the low-lying
states are high-field seekers. This limit is of most interest for our present
discussion. Within this high-field, pendular state model, the Stark shift,
$W$, is given by

\begin{equation}
W(v_{p},\lambda)/B=-\lambda + (v_{p}+1)(2\lambda)^{1/2},
\label{Eq:highFieldStark}
\end{equation}

\noindent
showing that the Stark shift becomes asymptotically linear in the electric
field.

We find it useful to define the effective
dipole moment, $\mu_{\mathrm{\it{eff}}}$, more generally as

\begin{equation}
\mu_{\mathrm{\it{eff}}}\scriptstyle(E)\displaystyle =
-\frac{\partial W}{\partial E},
\label{Eq:effdipole}
\end{equation}

\noindent
which converges with increasing electric field to its maximum value $\mu$,
reached when the body-fixed dipole moment is parallel to the external electric
field. From Eq.~(\ref{Eq:highFieldStark}) and (\ref{Eq:effdipole}) the
effective dipole moment is given, in the strong-field limit, by

\begin{equation}
\mu_{\mathrm{\it{eff}}}=\mu\left(1-\frac{v_{p}+1}{\sqrt{2\lambda}}\right).
\end{equation}

Note that the effective dipole moments of states having the same values of
$v_{p}$ converge once the strong-field criterion ($\lambda \gg 1$) is met. By
contrast, the convergence of all the $\mu_{\mathrm{\it{eff}}}$ to a single
value is a much slower one, scaling as $\lambda^{-1/2}$ and so requiring
$\lambda^{1/2} \gg 1$ for low values of $v_{p}$. In strong fields the effective
dipole moment varies little with applied field, and for small changes of the
field it can be approximated as a constant. This is a very useful approximation
in the context of an alternating gradient lens where the field is high and does
not vary greatly across the aperture of the lens.

Figure~\ref{Fig:Stark}(a) shows the Stark splitting and the effective dipole
moments for the $J$=1, $\Omega$= 1 level of the a$^{3}\Pi$ excited state of CO.
The electronic ground state of CO has a small dipole moment (0.1\,Debye) and
rotational levels in this state only experience a second order Stark effect in
realizable fields. By contrast, the metastable a$^{3}\Pi$ state of CO (lifetime
3.7\,ms) has a dipole moment of 1.37\,Debye (1\,Debye is equivalent to
0.0168\,cm$^{-1}$/kV/cm). Being a $\Pi$-state, all the rotational levels are
doubly degenerate. As the separation of the nuclear motion and the electronic
motion is not exact, this degeneracy is lifted and each rotational level is
split in zero electric field into two levels with opposite parity. For the
$J$=1, $\Omega$= 1 level this $\Lambda$-doublet splitting is $\Lambda$=394\,MHz.
The two $\Lambda$-doublet levels are coupled by an electric field, leading to
levels with a mixed parity that have non-zero space-fixed electric dipole
moment. The Stark shift of the two $\Lambda$-doublet levels in a small static
electric field of magnitude $E$ is found by diagonalizing the energy within a
single rotational manifold:

\begin{eqnarray}
W(E) = \pm
\sqrt{ \left( \frac{\Lambda}{2} \right)^{2}
+ \left( \mu E \frac{M\Omega}{J\left(J+1\right)}
\right)^{2}} \mp \frac{\Lambda}{2},
\label{eq:CO-Stark}
\end{eqnarray}

\noindent
where $J$ denotes the total angular momentum, while $\Omega$ and $M$ are the
projections of $J$ onto the body fixed and space fixed axes, respectively. At
higher electric fields, the Stark effect includes coupling to states of the
same $M$ but different $J$. Since low-lying states have many states of higher
$J$ above them, this coupling ultimately turns them all into high-field
seekers. For example, the uppermost level in Fig.~\ref{Fig:Stark}(a) is
weak-field seeking, but at fields above $\sim$400\,kV/cm it becomes high-field
seeking. The effective dipole moments for the lowest rotational levels of CO are
shown below the Stark-curves in Fig.~\ref{Fig:Stark}(a). Calculation of the
Stark shift in metastable CO is discussed in detail by Jongma et al.
\cite{Jongma:CPL270:304:1997}.

Figure~\ref{Fig:Stark}(b) shows the energy and effective dipole moments of YbF
in the $X^{2}\Sigma^{+}$ electronic ground state as a function of the electric
field strength \cite{Sauer:JCP105:7412:1996}. The states are labelled by the
rotational quantum number $N$ and its projection onto the electric field axis,
$M_{N}$. This Stark effect is caused by the mixing of rotational levels. The
ground rotational state is high-field seeking at all fields. Other states, such
as the $N=1, M_{N}=0$ state are low-field seeking at small electric fields but
become high-field seeking at larger field values.  For the $N=1, M_{N}=0$
state, the turning point occurs at an electric field of $\sim 5B/\mu$,
corresponding to only $\sim$18\,kV/cm for the heavy YbF molecule. For the fields
in Fig.~\ref{Fig:Stark}(b), the high-field condition $\lambda \gg 1$ is
satisfied, and one sees that the $\mu_{\mathrm{\it{eff}}}$ values converge for
states of the same $v_{p}=2J-|M|$, but different $M$. It is also evident in the
figure that convergence of the effective dipole moments to the single value,
$\mu$, is very slow, as discussed earlier.

In Fig.~\ref{Fig:Stark}(c) we show the Stark effect and effective
dipole moments in the lowest rotational states of benzonitrile,
calculated using experimentally determined constants
\cite{Borst:CPL350:485:2001}. Benzonitrile is an asymmetric top,
and therefore levels with the same $J$ are mixed by the electric
field as well as those having $\Delta J=\pm1$. As a molecule of
this size has rather small rotational constants, all rotational
levels become high-field seeking in relatively weak electric
fields. The jumps between dipole moment curves in
Fig.~\ref{Fig:Stark}(c) are caused by avoided crossings. Details
on the calculation of the Stark shifted energy levels in an
asymmetric top molecule can be found elsewhere
\cite{Bulthuis:JPCA104:1055:2000}.

\begin{table*}
\begin{center}
\input{Bethlem_Starktable}
\end{center}
\caption{\label{Tab:Stark} A selection of polar molecules with
their relevant properties for AG focusing and deceleration.
}\end{table*}

Table~\ref{Tab:Stark} gives the relevant properties for Stark
deceleration for a selection of polar molecules. These properties
are the Stark shift, effective dipole moment, rotational constants
and mass. Values are given at a field of 100\,kV/cm and are for
molecules in the electronic and rovibronic ground state (with the
exception of metastable CO). The number of electric field stages
required to bring molecules with a certain initial velocity to
rest depends on the ratio of their Stark shift to their mass. The
focusing properties of molecules flying at a certain velocity
depend on the ratio of their effective dipole moment to their
mass. One can see from the table that these molecules, though
widely different in mass, have similar ratios of effective dipole
moment to mass, and of Stark shift to mass, and so will be
focussed and decelerated similarly. The Stark shifts and effective
dipole moments are generally dependent on the specific quantum
state that the molecule is in. For many experiments one would like
to decelerate molecules in a variety of quantum states
simultaneously. This can be done when all the rotational states
have the same dependence on electric field, as is more or less the
case for the polyatomic aromatic molecules with small rotational
constants listed in the table. For complex molecules such as
tryptophan, the decelerator offers the intriguing possibility of
selecting a specific conformational isomer out of the various
conformers known to co-exist in a supersonic beam
\cite{Rizzo:JCP84:2534:1986}, as the individual conformers have
distinctly different values of $\mu_{\mathrm{\it{eff}}}$.

\section{\label{Sec:geometry}
Electrode Geometry}

\subsection{\label{SubSec:2D}
The field of an infinitely long lens}

Although the electrodes of a decelerator are in short segments along the beam
direction, the basic focusing properties are best elucidated by first
considering the case of long electrodes. In this section we discuss how to
design a set of electrodes that minimizes the aberrations of an AG lens. In an
aberration-free lens, molecules experience a harmonic interaction potential in
the transverse plane. As discussed in Sec.~\ref{Sec:GeneralPrinciples}, the
potential will focus along one direction and defocus along the other. For
molecules that experience a linear Stark shift the ideal form for the field
strength is also harmonic; $E(x,y) = E_{0} + \eta(x^{2} - y^{2})$. As we shall
see, this field cannot be realized but it is possible to produce a field that
is a good approximation to this ideal one. We follow a similar approach to that
given in \cite{Kalnins:RSI73:2557:2002}.

In a region devoid of charges the electric field can be derived from the
electrostatic potential $\Phi$ as $\vec{E}=-\vec{\nabla} \Phi$, with
$\nabla^{2} \Phi=0$. In 2D, $\Phi$ may be represented by a multipole expansion
as:

\begin{equation}
\begin{split}
\Phi(x,y) = \Phi_{0} \left[
\sum_{n=1}^{\infty}\frac{a_{n}}{n}
\left(\frac{r}{r_{0}}\right)^{n}\cos(n\theta) \right.\\
 + \left. \sum_{n=1}^{\infty}\frac{b_{n}}{n}
\left(\frac{r}{r_{0}}\right)^{n}\sin(n\theta) \right].
\end{split}
\label{Eq:multipole}
\end{equation}

\noindent 
Here $r=\sqrt{(x^{2} + y^{2})}$ and $\theta=\tan^{-1}\left(\frac{y}{x}\right)$
are the usual cylindrical coordinates. $a_{n}$ and $b_{n}$ are dimensionless
constants. $r_0$ and $\Phi_{0}$ are scaling factors that characterize the size
of the electrode structure and the applied voltages, respectively. The electric
field magnitude at the centre is given by $E_{0} =
(\Phi_{0}/r_{0})\sqrt{a_{1}^{2} + b_{1}^{2}}$. The $n=1$ terms in
Eq.~(\ref{Eq:multipole}) represent a constant electric field, while the $n=2$
and $n=3$ terms represent the familiar quadrupole and hexapole fields that have
been used extensively to focus molecules in low-field seeking states
\cite{Reuss:Book:1988}.

Equation~(\ref{Eq:multipole}) represents the most general form of the
electrostatic potential consistent with Laplace's equation. Now we choose the
coefficients to be suitable for making a good lens. We require the magnitude of
the electric field to be non-zero at the origin, and symmetric under reflection
in the $x$- and $y$-axes. To achieve this, we make $\Phi$ symmetric under
reflection in the $x$-axis and anti-symmetric under reflection in the $y$-axis
by setting all $b_{n}=0$ and retaining only the terms of odd $n$. Anticipating
the result that high-order terms only introduce undesirable non-linearities
into the force we choose to retain only $a_{1}$, $a_{3}$ and $a_{5}$. Hence:

\begin{equation}
\begin{split}
\Phi(x,y) =  \Phi_{0}\left(a_{1}\frac{x}{r_{0}} +
a_{3}\frac{\left(x^{3} - 3xy^{2}\right)}{3r^{3}_{0}} \right.\\
\left. + a_{5}\frac{\left(x^{5} - 10x^{3}y^{2} +
5xy^{4}\right)}{5r^{5}_{0}} \right).
\end{split}
\label{Eq:phi}
\end{equation}

\noindent From this potential, we obtain the electric field
magnitude, $E(x,y)=\sqrt{\left(\frac{\partial\Phi}{\partial
x}\right)^{2}+\left(\frac{\partial\Phi}{\partial y}\right)^{2}}$.
Throughout the region $r<r_{0}$ this can be expanded as a power
series in $a_{3}$ and $a_{5}$. For the case $a_{5} \ll a_{3} \ll
a_{1}$ we obtain

\begin{equation}
\begin{split}
E(x,y) =  E_{0} \left( 1 + \frac{a_{3}}{a_{1}}\frac{\left(x^{2} -
y^{2}\right)}{r^{2}_{0}} +    \right.\\ \left.
2\left(\left(\frac{a_{3}}{a_{1}}\right)^{2}
- 3\frac{a_{5}}{a_1}\right)
\frac{x^{2}y^{2}}{r^{4}_{0}}
 + \frac{a_{5}}{a_1} \frac{\left(x^{4} + y^{4}\right)}{r^{4}_{0}} +
\cdots \right).
\end{split}
\label{Eq:E}
\end{equation}

\noindent The first two terms have the desired form and dominate
the expansion. The other terms produce focusing aberrations. It
might appear advantageous to set $a_{5}/a_{1} =
(a_{3}/a_{1})^{2}/3$ so as to cancel the cross term. However, the
best policy is to minimize $a_{5}/a_{1}$
\cite{Kalnins:RSI73:2557:2002} because the $x^{4}+y^{4}$ term is
also a damaging aberration.

\setlength{\epsfxsize}{0.45\textwidth}
\begin{figure}
\centerline{\epsffile{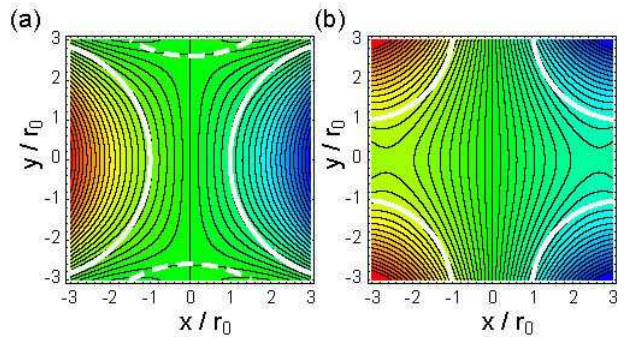}}
\protect
\vspace*{0.5cm}
\caption{\label{Fig:multipole} Electrostatic equipotentials of
equation (\ref{Eq:phi}) for the cases (a) $a_{3}/a_{1} = +1/7$,
$a_{5}=0$ and (b) $a_{3}/a_{1} = -1/7$, $a_{5}=0$. Red and blue
colourings correspond to positive and negative potentials
respectively. The white lines show electrode surfaces designed to
follow these contours.}\end{figure}

To produce these fields we need to choose electrodes whose
surfaces map onto the equipotentials. We are free to choose
$a_{3}/a_{1}$ either positive or negative and will discuss both
cases in turn. For example, Fig.~\ref{Fig:multipole}(a) shows
equipotentials for $a_{3}/a_{1} = +1/7$ and $a_{5}=0$. The choice
of $a_{3}$ is constrained by the condition $a_{3} \ll a_{1}$ while
remaining large enough to provide significant focusing. The solid
white lines show electrodes, chosen to be circular for ease of
construction, that match the equipotentials closely. They have
radii of $R=3r_{0}$ and are centred at $x=\pm4r_{0}$, leaving a
gap of $2r_{0}$.  Because these electrodes do not match the
equipotential exactly, higher order terms appear in the field.
From a fit to the numerically calculated electrostatic potential
we find for this geometry $a_{3}/a_{1}=0.143$ and
$a_{5}/a_{3}=0.143$. It is noted that this two-rod field can be
solved analytically and that $a_{3}/a_{1}= \left(r_{0}/R\right)/
\left( 2 + r_{0}/R\right)$ and $a_{5}/a_{3}$ = $a_{3}/a_{1}$, in
agreement with our fit.

\setlength{\epsfxsize}{0.45\textwidth}
\begin{figure}
\centerline{\epsffile{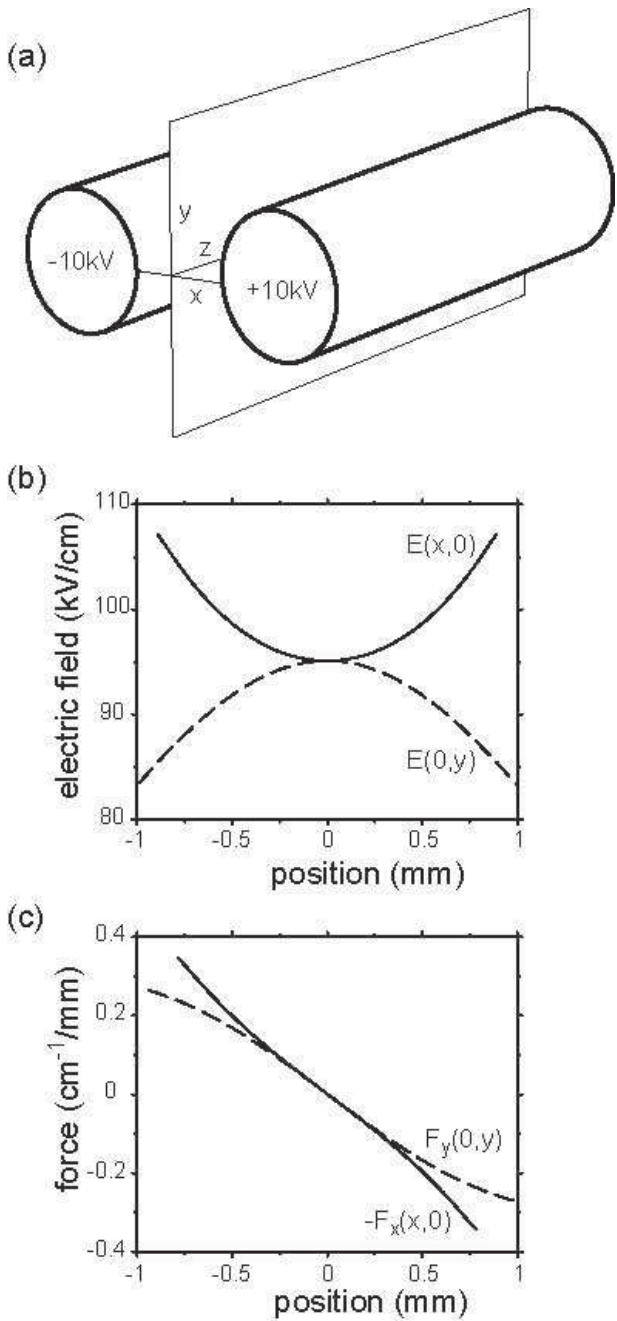}}
\protect
\vspace*{0.5cm}
\caption{\label{Fig:dipolelens} The case of a two-rod lens. (a)
Schematic view of the lens formed by two 6\,mm diameter rods spaced
2\,mm apart with a potential difference of 20\,kV. (b) The electric
field strength versus displacement along the $x$-axis (solid line)
and the $y$-axis (dashed line). (c) Forces on a CO molecule in the
high-field-seeking component of the a$^{3}\Pi$, $J$=1, $\Omega$=1
level. Dashed line: focusing force, $F_{y}(0,y)$. Solid line:
defocusing force $F_{x}(x,0)$. The sign of this force has been
reversed for ease of comparison. The gradient of both lines near
the origin is $k$=0.37\,cm$^{-1}$/mm$^{2}$. }\end{figure}

The two charged rods are schematically depicted in
Fig.~\ref{Fig:dipolelens}(a), while Fig.~\ref{Fig:dipolelens}(b)
shows the magnitude of the electric field they produce as a
function of distance along the $x$-axis (solid line) and $y$-axis
(dashed line). A high-field seeker will be defocussed along $x$
and focussed along $y$. Fig.~\ref{Fig:dipolelens}(c) shows the
corresponding forces exerted on CO molecules in the high field
seeking component of the a$^{3}\Pi$, $J$=1, $\Omega$=1 level. The
dashed line is the restoring force generated by a displacement
along $y$, and is seen to be roughly linear. The solid line
indicates the defocusing force along $x$. The sign of this force
has been reversed so that the two can be compared directly. The
two have equal gradients near the origin. Further away from the
origin, the non-linearity due to $a_{5}$ acts to strengthen the
defocusing power whereas the focusing is weakened. We will see
later that this difference reduces the acceptance of an
alternating gradient decelerator.

The rather large value of $a_{5}/a_{3}$ in this two-rod case can be reduced by
adding two grounded electrodes tangential to the $\Phi = 0$ equipotentials at
$y=\pm 2.65r_{0}$. These are shown dashed in Fig.~\ref{Fig:multipole}(a), where
for simplicity we have given the new rods the same radius $R$. In this case the
coefficients become $a_{3}/a_{1}=0.157$ and $a_{5}/a_{3}=0.070$. At the expense
of a slightly less ideal field one can position the four identical electrodes
at the corners of a square. This has the advantage that one is free to choose
in which plane the field focuses or defocuses by simply switching the voltages
\cite{Anderson}. Using electrodes of radius $r_{0}$ with their centers
placed on the corner of a square of side 3$r_{0}$ yields $a_{3}/a_{1}=0.59$
and $a_{5}/a_{3}=0.056$. The rather large value of $a_{3}/a_{1}$ introduces
higher order terms in the field, even though $a_{5}/a_{3}$ is quite small. A
disadvantage of this field geometry is that the electric field strength on the
beam axis is only half that on the electrodes. This makes the configuration
less suited for use in a decelerator as the energy removed per stage is
proportional to the central field. The geometry is useful for guiding
molecules, as was recently demonstrated by Junglen et al.
\cite{Junglen:PRL92:223001:2004}.

We turn now to the case of negative $a_{3}/a_{1}$ illustrated in
Fig.~\ref{Fig:multipole}(b) where we have chosen
$a_{3}/a_{1}=-1/7$ and $a_{5}=0$. This is well approximated by
electrodes of radius $R=2.3r_{0}$, with a minimum gap of $2r_{0}$
as shown by the solid white lines of Fig.~\ref{Fig:multipole}(b).
The precise field produced by these electrodes has
$a_{3}/a_{1}=-0.139$ and $a_{5}/a_{3}=-0.014$. This geometry
compares very favourably to the cases considered in
Fig.~\ref{Fig:multipole}(a) with regard to minimizing $a_{5}$ and
hence the lens aberrations. As this geometry is symmetric, we can
again reverse the focus and defocus directions very easily by
interchanging the potentials on the top-right and bottom-left
electrodes of Fig.~\ref{Fig:multipole}(b). The field at the centre
is 41\% of the maximum, which is disadvantageous for a
decelerator. This geometry was discussed by L\"ubbert et al.
\cite{Lubbert2} in the context of focusing ICl molecules.

\setlength{\epsfxsize}{0.45\textwidth}
\begin{figure}
\centerline{\epsffile{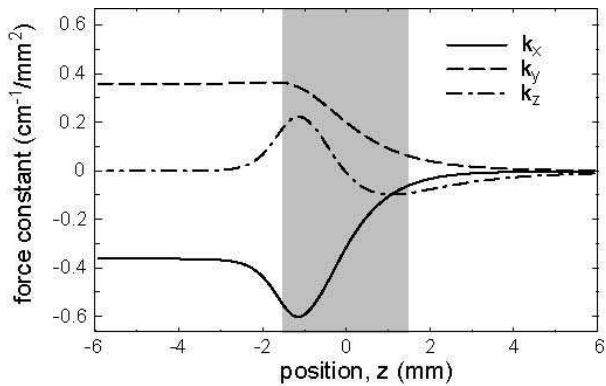}}
\protect
\vspace*{0.5cm}
\caption{\label{Fig:endeffects} The `force constants' for CO
(a$^{3}\Pi$, $J$=1, $\Omega$=1, $M\Omega$=+1) along the molecular
beam axis and near the exit of the lens shown in
Fig.~\ref{Fig:ExperimentLayout}(b) and
Fig.~\ref{Fig:dipolelens}(a). Solid line: defocussing constant
$k_{x}=-\partial F_{x}/\partial x$. Dashed line: focussing
constant $k_{y}=-\partial F_{y}/\partial y$. Dash-dotted line:
axial gradient $k_{z}=-\partial F_{z}/\partial z$. The three are
linked by Eq.~(\ref{Eq:divFOnAxis}). The gray shaded area
indicates the region of the lens's hemispherical end.
}\end{figure}

\subsection{\label{SubSec:endeffects}
End effects}

Until now we have assumed the electrodes to be infinitely long. We
now discuss the influence of end effects.
Eq.~(\ref{Eq:divergenceOfForce}) gives the divergence of the force
on a molecule with linear Stark shift. If we restrict our
attention to the axis of our beamline, this equation is greatly
simplified. Consider, for example, the pair of electrodes shown in
Fig.~\ref{Fig:dipolelens}. The electrostatic potential is
symmetric about the $x$-$z$-plane and the $y$-$z$-plane is one of
antisymmetry with $\Phi = 0$ everywhere on this plane. All the
electrode geometries considered here have this property. On the
beamline, the intersection of these two planes, it follows that
all the derivatives in equation (\ref{Eq:divergenceOfForce}) are
zero apart from $\partial \Phi /
\partial x$ and $\partial^{2}\Phi / \partial z \partial x$. Since
$-\partial \Phi / \partial x$ is the only non-zero electric field
component, its magnitude is the total electric field strength $E$.
Thus, on axis,

\begin{equation}
\vec{\nabla}\cdot \vec{F} =
\frac{\mu_{\mathrm{\it{eff}}}}{E}\left(\frac{\partial E}{\partial
z}\right)^{2}. \label{Eq:divFOnAxis}
\end{equation}

\noindent Inside the lens $\partial E/\partial z = 0$ and
$\partial F_{z}/\partial z = 0$ and it follows that  the spring
constants in the two transverse directions $k_{x} = -\partial
F_{x}/\partial x, k_{y} = -\partial F_{y}/\partial y$ are equal
and opposite, as shown in Fig.~\ref{Fig:dipolelens}(c). The
equality $k_{x}=-k_{y}$ means that the focusing and defocusing
powers are equal. In the fringe field  of the lens $\partial
E/\partial z \neq 0$ and $\partial F_{z}/\partial z \neq 0$ and we
find that

\begin{align}
k_{x}+k_{y} &=
-\mu_{\mathrm{\it{eff}}}\left(\frac{1}{E}\left(\frac{\partial
E}{\partial z}\right)^{2}-\frac{\partial^{2}E}{\partial
z^{2}}\right) \nonumber \\
&= \mu_{\mathrm{\it{eff}}} E \frac{\partial}{\partial
z}\left(\frac{1}{E} \frac{\partial E}{\partial z}\right).
\label{Eq:kxPlusky}
\end{align}

\noindent 
Due to the inhomogeneity of the electric field along $z$, the
defocusing force becomes larger than the focusing force near the exit of the
lens, whereas the focusing force is larger than the defocusing force further
away from the lens. This is illustrated in Fig.~\ref{Fig:endeffects}, that
shows the three force constants $k_{x}, k_{y}$ and $k_{z}$ for the high-field
seeking CO molecules as they approach the end of a lens formed by two rods with
hemispherical ends. In this figure, the origin of $z$ is at the point of
inflection ($\partial^{2} E / \partial z^{2}$=0) and the region of the
hemispherical ends is indicated by the grey shaded area. We begin with the left
hand side of the figure where the end-effects are negligible. Here, $k_{z}=0$
and therefore $k_{x}=-k_{y}=0.37$\,cm$^{-1}$/mm$^{2}$ the same as in
Fig.~\ref{Fig:dipolelens}. As the molecules approach the exit of the lens, they
experience a decelerating force which can be seen in the figure as a positive
$k_{z}$. According to Eq.~(\ref{Eq:divFOnAxis}), this is accompanied by a
corresponding decrease in the sum $k_{x}+k_{y}$. We see this as a strengthening
of the defocusing constant $k_{x}$ near the exit, which becomes nearly twice as
strong as the focusing constant. As we will see in
Sec.~\ref{subSec:transverse}, this is another mechanism, in addition to the
aberrations discussed in Sec.~\ref{SubSec:2D}, that significantly reduces the
transmission of an AG decelerator. A more gradual termination of the rods, e.g.
a prolate spheroid replacing the hemisphere, reduces both the first and second
derivatives of $E$ with respect to $z$ and so reduces end-effects in accordance
with Eq.~(\ref{Eq:kxPlusky}). In comparison with the two-rod configuration, the
four-rod arrangement of Fig.~\ref{Fig:multipole}(b) is also found to have a
more favourable field at the exit of the lens. 

\section{\label{Sec:AG}
Motion of the molecules through the decelerator}

In this section we investigate the motion of molecules through the
decelerator and discuss criteria for optimizing the transmission.
The first part of this section deals with transverse stability,
and the second part with longitudinal stability. This division is
based on the assumption that the transverse and longitudinal
motions can be treated independently. This is an approximation
whose validity we discuss at the end of the section.

\subsection{\label{subSec:transverse}
Transverse motion}

In describing the transverse motion, we start by assuming that the molecules
experience a linear force that focuses them along one direction and defocuses
them along the other. The orientation of successive lenses alternates. The
lenses have lengths $L$ and are separated by drift regions of length $S$ where
the molecules experience no force. For molecules moving with a constant
velocity, $v_{z}$, along the molecular beam axis, the equation of motion in a
lens can be written as $\partial^2x/\partial z^2 \pm \kappa^{2}x = 0$, where
the plus sign applies in a focusing lens, and the minus sign applies in a
defocusing lens. The number of oscillations per unit length inside a focusing
lens is $\kappa/2\pi$ and is related to the force constant $k$ by $\kappa =
\sqrt{\vert k \vert /mv^{2}_{z}}$. We also define the angular oscillation
frequency $\Omega$ which, for the linear Stark effect reads

\begin{equation}
\Omega=\sqrt\frac{|k|}{m}=\sqrt{\frac{\mu_{\mathrm{\it{eff}}}}{m}\frac{2E_{0}a_{3}}{r_{0}^{2}}}.
\label{Eq:omega}
\end{equation}

\noindent For a molecule with initial position $x(z_{0})$ and
velocity $v(z_{0})$, the equation of motion can be written as

\begin{eqnarray}
\left( \begin{array}{cc}
x(z)\\ v_{x}(z)
\end{array} \right)  &=&
M(z\vert z_{0})
\left(\begin{array}{cc}
x(z_{0})\\ v_{x}(z_{0})
\end{array} \right).
\end{eqnarray}

\noindent
The transfer matrix $M(z\vert z_{0})$ is then given by

\begin{eqnarray}
M(z\vert z_{0}) = \left\{
\begin{array}{ll}
{\left( \begin{array}{cc}
\cos{\kappa l} &
\frac{1}{\Omega}\sin{\kappa l} \mbox{~}\\
-\Omega \sin{\kappa l} &
\cos{\kappa l} \mbox{~}
\end{array} \right)}& F\mbox{:~~focusing lens}\\
{\left( \begin{array}{cc}
 1 & l/v_{z}\\
 0 & 1  \end{array} \right)}& O\mbox{:~~drift space}\\
{\left( \begin{array}{cc}
\cosh{\kappa l} &
\frac{1}{\Omega}\sinh{\kappa l} \mbox{~}\\
\Omega \sinh{\kappa l} &
\cosh{\kappa l} \mbox{~}
\end{array} \right)}& D\mbox{:~~defocusing lens}
\end{array}
\right.
\label{Eq:transferMatrices}
\end{eqnarray}

\noindent
where $l= z-z_{0}$. The transfer matrix is written as $F$ in a
focusing lens, as $D$ in a defocusing lens and as $O$ in a drift region.

The transfer matrix for any interval made up of subintervals is just the
product of the transfer matrices of the subintervals:

\begin{eqnarray}
M(z_{2}\vert z_{0}) = M(z_{2}\vert z_{1})M(z_{1}\vert z_{0}).
\end{eqnarray}

\noindent A single repeating unit of the alternating gradient array has the
transfer matrix $F(L).O(S).D(L).O(S)$. We have written the lengths $L$ and $S$
explicitly here, but will usually drop them. The transfer matrix for an array
of $N$ such units is $M=(FODO)^{N}$. Alternatively, it can be useful to
introduce frequent deceleration sections into longer lenses using a
configuration $M=(FO)^{n}(DO)^{n}$.  This structure with $n=3$ is used in the
decelerator that we present in Sec.~\ref{Sec:experimental}. In order for
molecules to have stable trajectories through any such array it is necessary
that all the elements of the transfer matrix remain bounded when $N$ increases
indefinitely. This is the case when $-1 < \frac{1}{2} Tr (M) < +1$ (see, for
example, \cite{Lee:Book}).

It is useful to parameterize the transfer matrix of one repetitive unit with
length $l_{\mathrm{\it{cell}}}$ as \cite{Courant:AnnPhys3:1:1958}

\begin{eqnarray}
M(z+l_{\mathrm{\it{cell}}}|z)= \left( \begin{array}{cc} \cos{\Phi}
+ \alpha \sin{\Phi}
& \beta \sin{\Phi} \\
-\gamma \sin{\Phi} &
\cos{\Phi} - \alpha \sin{\Phi}
\end{array} \right),
\label{Eq:Courant-Snyder}
\end{eqnarray}

\noindent where $\alpha(z)$, $\beta(z)$ and $\gamma(z)$ are
z-dependent parameters with periodicity equal to that of the
lattice and are known as the Courant-Snyder parameters. $\Phi$ is
known as the phase-advance per cell. Note that $\beta(z)$ and
$\gamma(z)$ are expressed in seconds and 1/seconds, respectively,
rather then in meters and 1/meters as is customary in the charged
particle accelerator literature. This follows from our use of
($x,v_{x}$) as state variables, rather than ($x,v_{x}/v_{z}$).

The Courant-Snyder parameters and the phase advance are related to one another:

\begin{subequations}
\begin{equation}
\label{Eq:alpha} \alpha(z) = -\frac{v_{z}}{2}\frac{d
\beta(z)}{dz},
\end{equation}
\begin{equation}
\label{Eq:gamma} \gamma(z) = \frac{1+\alpha^{2}(z)}{\beta(z)},
\end{equation}
\begin{equation}
\label{Eq:phaseAdvance}
\Phi =
\frac{1}{v_{z}}\int_{z}^{z+l_\mathrm{\it{cell}}}{\frac{1}{\beta(z')}\,dz'}.
\end{equation}
\end{subequations}

\noindent Equation~(\ref{Eq:gamma}) ensures that the matrix has
unity determinant. When expressed in this form, the transfer
matrix acquires an extremely useful property, namely that the
matrix describing $N$ lattice units is identical to the matrix for
a single unit, but with $\Phi$ replaced by $N\Phi$.
Equation~(\ref{Eq:phaseAdvance}) shows that $\Phi$ is independent
of $z$, since the integral is taken over one complete period of
the periodic function $\beta$. Note that the stability criterion
becomes $-1 < \cos{\Phi} < +1$ and so is satisfied if $\Phi$ is
real.

The trajectory of a molecule moving through the ideal lattice is given by

\begin{equation}
x(z)=\sqrt{\beta(z)\epsilon_{i}}\cos(\phi(z)+\delta_{i}),
\label{Eq:moleculeTrajectory}
\end{equation}

\noindent where $\epsilon_{i}$ and $\delta_{i}$ define the initial
conditions of this particular molecule, and $\phi(z)$ is a
z-dependent phase given by
$\phi(z)=1/v_{z}\int_{0}^{z}{1/\beta(z')\,dz'}$.
Equation~(\ref{Eq:moleculeTrajectory}) shows that the motion is a
product of two periodic functions, the first of wavelength
$l_{\mathrm{\it{cell}}}$ and the second of wavelength $2\pi
l_{\mathrm{\it{cell}}}/\Phi$. When $\Phi \ll 2\pi$, the first
motion has a short wavelength and is known as the micromotion,
while the second has a much longer wavelength and is called the
macromotion. This motion is identical to that of an ion in an rf
trap \cite{Paul:RMP62:531:1990}. From $x(z)$ and the relationships
that hold between the Courant-Snyder parameters, it can be shown
that

\begin{equation}
\gamma(z) x^{2} + 2\alpha(z) xv_{x} + \beta(z)
v^{2}_{x} = \epsilon_{i}.
\label{Eq:ellipse}
\end{equation}

This equation defines an ellipse in the phase-space whose coordinates are $x$
and $v_{x}$. The shape of the ellipse evolves periodically with $z$, but always
has the same area $\pi\epsilon_{i}$. A set of molecules having many different
values of $\delta_{i}$ but the same value of $\epsilon_{i}$ will all lie on the
same ellipse. Furthermore, a distribution of molecules with all possible values
of $\delta_{i}$ and all values of $\epsilon_{i}$ in the range
$0<\epsilon_{i}<\epsilon$, will all lie inside the ellipse characterized by
$\epsilon$. Again, the shape of this ellipse evolves periodically, but its area
is a constant, $\pi\epsilon$. The value of $\epsilon$ defines the size of the
beam in phase-space, and is called the emittance of the beam.
Equation~(\ref{Eq:moleculeTrajectory}) tells us that the transverse
displacements of a set of molecules lie within a beam envelope given by the
periodic function $\pm\sqrt{\beta(z)\epsilon}$. The velocity spread lies within
a beam envelope given by $\pm\sqrt{\gamma(z)\epsilon}$.

\setlength{\epsfxsize}{0.45\textwidth}
\begin{figure}
\centerline{\epsffile{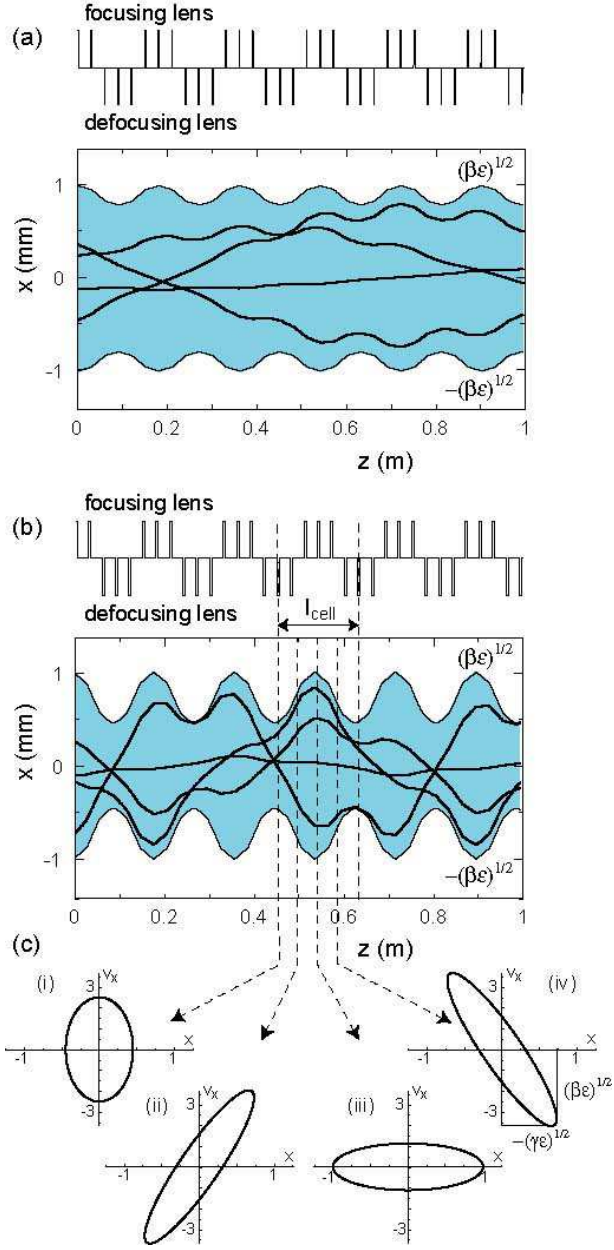}}
\protect
\vspace*{0.5cm}
\caption{\label{Fig:trajectories}
(a) Trajectories of metastable CO molecules flying with a forward speed of
315\,m/s through a $(FO)^{3}(DO)^{3}$ array. The parameters are
$\kappa$=38.7\,m$^{-1}$, $L$=2\,mm and $S$=28\,mm, corresponding to $\Phi=\pi/6$.
The shaded area shows the beam envelope bounded by $\pm\sqrt{\beta \epsilon}$,
for a constant aperture of $d$=2\,mm. (b) As (a), but with $L$=6\,mm and
$S$=24\,mm corresponding to $\Phi=\pi/2$. (c) Phase-space area occupied by the
beam at four positions in the unit cell with the position spread in mm and the
velocity in m/s. (i) Centre of the defocusing set, (ii) between the defocusing
and focusing sets, (iii) centre of the focusing set and (iv) between the
focusing and defocusing sets.
}\end{figure}

In Fig.~\ref{Fig:trajectories}(a), some trajectories are plotted for metastable
CO molecules travelling at 315\,m/s through an alternating gradient array of
type $(FO)^{3}(DO)^{3}$. The parameters of the array are
$\kappa=38.7$\,m$^{-1}$, $L=2$\,mm and $S=28$\,mm. Here, the phase-advance is
$\pi/6$, and the micromotion of wavelength $l_{\mathrm{\it{cell}}}$ is
superimposed on a macromotion whose wavelength is 12\,$l_{\mathrm{\it{cell}}}$.
The shaded area of the figure shows the envelope of the transmitted molecular
beam as it passes through the array. For such small values of the
phase-advance, there is only a small difference between the maximum and minimum
sizes of the beam envelope. As the phase-advance increases, the modulation of
the beam envelope increases. This is demonstrated in
Fig.~\ref{Fig:trajectories}(b) which shows trajectories and beam envelope for
the same value of $\kappa$ but with $L$ increased to $6$\,mm and $S$ decreased
to $24$\,mm. The phase advance is now $\pi/2$, meaning that molecules return to
their starting point after 4\,$l_{\mathrm{\it{cell}}}$.
Figure~\ref{Fig:trajectories}(c) shows the phase-space distribution of the beam
at four positions within the unit cell. In graph (i), the molecules are at the
centre of the defocusing triplet. Here, the transverse size of the beam is at
its minimum. The beam is diverging as it enters the focusing lenses (ii), and
reaches its maximum size at the centre of the focusing triplet (iii). Graph
(iv) shows that the beam is converging when it enters the defocusing lenses.
The fact that the transverse size of the beam is larger in the focusing lenses
than in the defocusing lenses, and that the forces are proportional to the
off-axis displacements, accounts for the stability of the array. Since the
stability relies upon the motion itself, it is commonly referred to as
`dynamic' stability.

\setlength{\epsfxsize}{0.45\textwidth}
\begin{figure}
\centerline{\epsffile{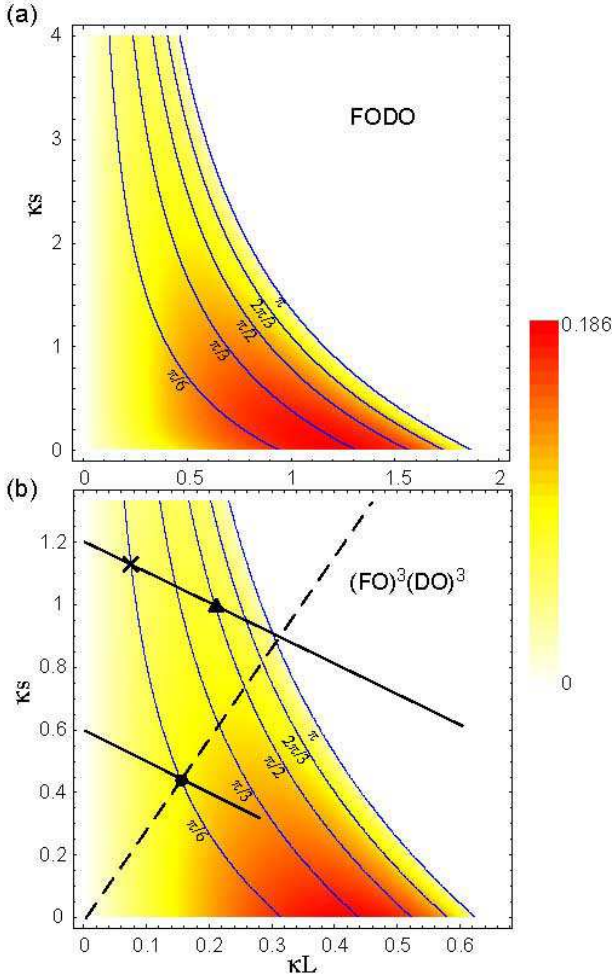}}
\protect
\vspace*{0.5cm}
\caption{\label{Fig:acceptance}
Acceptance in one transverse direction of an
infinitely long alternating gradient array, as a function of $\kappa L$ and
$\kappa S$. The acceptance is in units of $d^{2}\Omega$. Also shown are some
contours of $\Phi$ which define the stability region. (a) A $FODO$ array. (b) A
$(FO)^{3}(DO)^{3}$ array. Straight solid and dashed lines, cross, triangle
and dot are all referred to in the text.
}\end{figure}

We now calculate the transverse acceptance of the AG array. If we suppose that
the array of lenses has a uniform aperture $d$ throughout its length, then the
beam whose emittance is $\epsilon$ will be transmitted without loss provided
that the envelope fits inside the aperture, i.e. provided that
$\sqrt{\beta(z)\epsilon} < d/2$ everywhere in the array. The transverse
acceptance is the phase-space area occupied by the beam of largest emittance
consistent with this criterion. This area is $\pi d^2/(4\beta_{max})$.
From Fig.~\ref{Fig:trajectories} we see that $\beta$ is always a maximum at the
centre of a focusing lens. To calculate the transverse acceptance of a lattice,
we simply find the value of $\beta$ at this position using
Eq.~(\ref{Eq:Courant-Snyder}) with $M=F(L/2) \cdot O(S)\cdot D(L)\cdot
O(S)\cdot F(L/2)$. From this we find that $\beta$ is a dimensionless number
divided by $\Omega$ and so the acceptance is a multiple of $d^{2}\Omega$.
$d$ and $\Omega$ are the natural scaling parameters of the problem, and
trajectories are invariant when plotted in $(x/d,v_{x}/(d\Omega))$ space.

In Fig.~\ref{Fig:acceptance}(a) and Fig.~\ref{Fig:acceptance}(b)
we show the transverse acceptance calculated for lattices of
$FODO$ and $(FO)^{3}(DO)^{3}$ cells, respectively. The acceptance
is plotted as a function of the two dimensionless parameters that
define the lattice, $\kappa L$ and $\kappa S$. The acceptance (in
either transverse direction) is given in units of $d^{2}\Omega$.
One sees that the highest transverse acceptance is
$0.186d^{2}\Omega$ and is obtained when $\kappa L \sim 1$ and $S
\ll L$. By contrast, a single, infinitely long focusing lens has
an acceptance of $(\pi/4)d^{2}\Omega$, which is over four times
larger. The figure also shows some contours of $\Phi$, filling
the region of stability bounded by the $\cos\Phi=\pm 1$ contours.
The $\cos\Phi=+1$ contour corresponds to the vertical line $\kappa
L=0$.

In the experiments discussed in Sec.~\ref{Sec:experimental}, the
physical structure is fixed at $L+S=30$\,mm, but the effective
length of the lenses can be varied by adjusting the high-voltage
switch-on time (see Fig.~\ref{Fig:ExperimentLayout}(b)). Thus, the
possible operating conditions lie on a straight line. Two such
lines are shown in Fig.~\ref{Fig:acceptance}(b), indicating the
operating conditions for the experiments where metastable CO was
used at forward speeds of 630\,m/s and 315\,m/s. The cross and
triangle placed on Fig.~\ref{Fig:acceptance}(b) correspond to the
settings used to calculate the  trajectories in
Fig.~\ref{Fig:trajectories}(a) and Fig.~\ref{Fig:trajectories}(b),
with small and large phase advance respectively.

It is worth considering the scaling behaviour of the transverse
acceptance with aperture, $d$. The acceptance along each direction
scales as $d^{2}\Omega$, and $\Omega$ scales as
$E_{0}^{1/2}d^{-1}$ (Eq.~(\ref{Eq:omega})). It is natural to
operate the decelerator at the maximum field that can be achieved,
which is determined by the breakdown field. If the transverse
scale of the lenses is increased, with $E_{0}$ held constant by
corresponding increases in the applied voltages, the acceptance
along each direction is linear in $d$, implying that one should
make the aperture as large as possible. There is, however, a
practical upper bound $V_{max}$ on the applied voltages. Once this
value of $V_{max}$ is reached, $E_{0}$ scales as $d^{-1}$ and the
acceptance in each transverse direction is proportional to
$d^{1/2}$. Although the acceptance continues to increase with $d$,
the decreasing value of $E_{0}$ results in an undesirable decrease
in the energy loss per deceleration stage. Furthermore, increases
in the acceptance cease to be useful once the transverse emittance
of the beam is fully contained within the transverse acceptance of
the decelerator.

\setlength{\epsfxsize}{0.45\textwidth}
\begin{figure}
\centerline{\epsffile{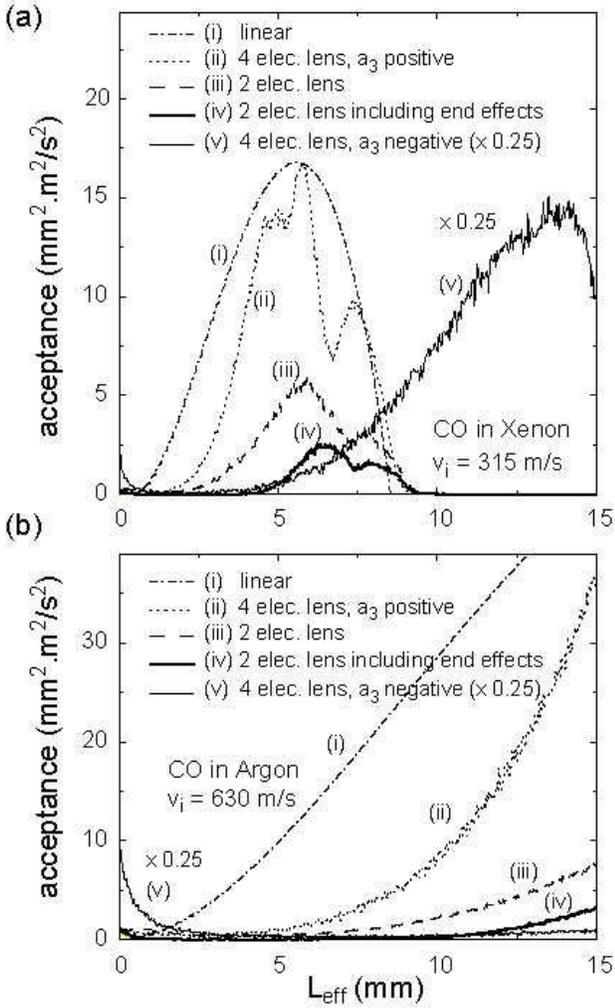}}
\protect
\vspace*{0.1cm}
\caption{\label{Fig:online}
The acceptance along the solid lines indicated in Fig.~\ref{Fig:acceptance} for
(a) CO seeded in Xenon ($v_z$=315\,m/s) and (b) CO seeded in Argon
($v_z$=630\,m/s). Lenses with four different electrode geometries have been used,
as indicated in the legend. For comparison, the transmission of perfectly
linear lenses with a minimum aperture of 2.1\,mm is also shown.
}\end{figure}

We have seen that transverse stability depends on maintaining
coherence between the oscillation of the molecules and the
structure of the array. This makes the alternating gradient
focussing particularly sensitive to deviations from the ideal,
such as nonlinear terms in the force, end effects, and
misalignments of the lens array. We now discuss the impact of each
of these on the transverse acceptance. Fig.~\ref{Fig:online} shows
the two-dimensional transverse acceptance of an $(FO)^{3}(DO)^{3}$
array calculated for the various lens geometries discussed in
Sec.~\ref{Sec:geometry}. The lens-lens spacing, $L+S$, is fixed at
30\,mm, and the operating conditions correspond, for metastable
CO, to the two straight, solid lines drawn on
Fig.~\ref{Fig:acceptance}(b). Figure \ref{Fig:online}(a) is
calculated for CO molecules with a forward speed of 315\,m/s,
while Fig.~\ref{Fig:online}(b) is for a speed of 630\,m/s. The
calculation uses the known Stark shift of metastable
CO~\cite{Jongma:CPL270:304:1997} and the electric fields obtained
from SIMION \cite{SIMION}. The trajectories of, typically,
5\,10$^{5}$ molecules with random initial positions and velocities
are traced through a 96-lens array by numerical integration. Line
(i) shows the acceptance obtained for a set of perfect linear
lenses with $E_{0}=95$\,kV/cm, $a_{3}/a_{1}=1/7$ and
$r_{0}=1$\,mm. The lens aperture, $d$, is taken to be 2.1\,mm for
reasons that will become clear shortly. Since this line is for
perfect lenses, it sets the scale for the forthcoming cases. Line
(ii) shows the acceptance obtained when the force has a small
non-linearity, corresponding to the real field of the four lens
geometry of Fig.~\ref{Fig:multipole}(a), with the two high voltage
electrodes held at $\pm$10\,kV and the other two grounded. Once
again, $r_{0}$ is 1\,mm and this gives $E_{0}$=95\,kV/cm.
End-effects are not considered. The curve displays structure that
is absent in the ideal case, and the region of high acceptance is
seen to be narrower. We chose the value of $d$ in the ideal case
(curve (i)) so as to give the same maximum acceptance. This
suggests the definition of an effective aperture,
$d_{\mathrm{\it{eff}}}$, 2.1\,mm in this case. It is interesting
to note that this effective aperture is slightly larger than the
real 2\,mm gap between the two electrodes at high voltage. This
occurs because the beam envelope is smaller in the defocusing
direction than in the focusing direction, and the smaller gap is
in the defocusing direction. The 2D transverse acceptance scales
as $d_{\mathrm{\it{eff}}}^{4}$. We next increase the size of the
non-linear contributions to the force, by removing the two
grounded electrodes of Fig.~\ref{Fig:multipole}(a). As discussed
in Sec.~\ref{Sec:geometry}, this approximately doubles the ratio
$a_{5}/a_{3}$. All other parameters are kept constant, and
end-effects are not yet considered. The acceptance in this case is
given by line (iii). We find the impact of the non-linearities to
be very detrimental indeed. The effective aperture is reduced to
1.6\,mm. As shown by line (iv), a further reduction in acceptance
occurs when we introduce the fringe-field aberrations at the
entrance and exit of each lens. Here, the electrodes of the
two-rod lens have hemispherical ends of radius 3\,mm as outlined
in Sec.~\ref{SubSec:endeffects}. The effective aperture for this
case is 1.3\,mm. We have also considered the four-electrode
geometry of Fig.~\ref{Fig:multipole}(b), with the electrodes at
$\pm$10\,kV and $r_{0}=1$\,mm, giving $E_{0}=47$\,kV/cm. As
discussed in Sec.~\ref{Sec:geometry}, this geometry results in
very small non-linearities, the ratio $a_{5}/a_{3}$ being a factor
of 10 smaller than in the two-rod case. The calculated acceptance
is shown by line (v) and is divided by a factor 4 for ease of
comparison. As $E_{0}$, and hence $\Omega$ is smaller in this
case, the maximum acceptance is shifted to higher values of $L$.
The effective aperture for this configuration is 2.9\,mm making it
by far the most effective configuration considered. Note that this
effectiveness is due to the small value of $a_{5}/a_{3}$ and not
due to the sign of $a_{5}/a_{3}$; indeed, calculations show that
if we reverse the sign the acceptance is the same. As mentioned in
Sec.~\ref{Sec:geometry}, a disadvantage of this geometry is that
the field on axis is only half that of the two-rod geometry, for
the same maximum electric field. A decelerator composed of these
4-rod lenses would therefore require twice as many deceleration
stages as one composed of 2-rod lenses. The much improved
acceptance comes at the cost of increased decelerator length.

\setlength{\epsfxsize}{0.45\textwidth}
\begin{figure}
\centerline{\epsffile{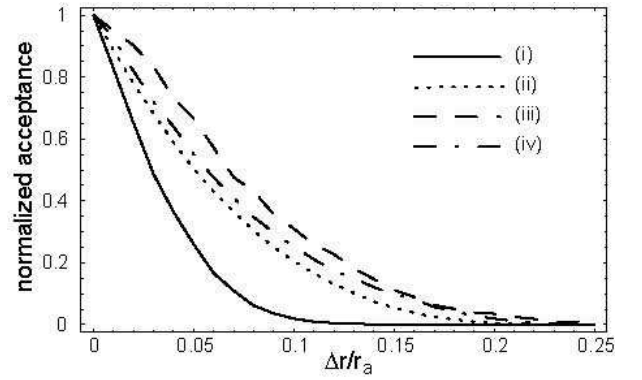}}
\protect
\vspace*{0.5cm}
\caption{\label{Fig:misalignment}
The calculated acceptance for an array of 96 ideal lenses, as a function of the
degree of misalignment (defined in the text). Random (i) and constant (ii)
misalignments of a $FODO$ array with $\kappa$=50\,m$^{-1}$, L=S=15\,mm.
Random (iii) and constant (iv) misalignments of a $(FO)^{3}(DO)^{3}$
array with $\kappa$=38.7\,m$^{-1}$, L=6\,mm and S=24\,mm.
}\end{figure}

Finally in this section, we discuss how the inevitable misalignments of a real
machine affect the transverse acceptance. We consider two types of
misalignment. In the first type, referred to as `random', the centre of each
lens is displaced horizontally and vertically from the axis by amounts chosen
at random from normal distributions with full-width at half-maximum $\Delta r$.
In the second type, referred to as `constant', lenses that focus in the
horizontal direction are perfectly aligned to one another and define the axis,
while the lenses that focus in the vertical direction are all displaced from
the axis by an amount $\Delta r$ in both transverse directions. These types of
misalignment tend to occur naturally in the construction of the decelerator.
For example, in our decelerators, each electrode is mounted into one of four
common bars to which the high voltages are applied. The degree of random
misalignment is determined by the machining precision and the construction
technique. Misalignment of the four bars relative to each other results in a
constant displacement of the horizontal lenses from the vertical lenses.

We have calculated how the transverse acceptance of the decelerator diminishes
as the degree of misalignment increases. Sensitivity to misalignments is found
to depend somewhat on the values of $\kappa$, $L$ and $S$. Some representative
cases are shown in Fig.~\ref{Fig:misalignment}. This figure gives the 2D
transverse acceptance for an alternating gradient array of 96 ideal lenses
(linear force, no end effects) as a function of the degree of misalignment,
$\Delta r/ r_{a}$, $r_{a}$ being the radius of the circular aperture defined by
the lenses. Each line in the figure has been normalized to the acceptance
obtained for perfect alignment. Line (i) gives the result in the case of random
misalignments in a $FODO$ array with $\kappa$=50\,m$^{-1}$, $L$=15\,mm and
$S$=15\,mm. One sees that random misalignments of $\Delta r \sim 0.03r_{a}$ are
sufficient to reduce the acceptance by 50\%. Line (ii) plots the effect of a
constant misalignment for the same parameters, showing this to be a less severe
misalignment in this case. A 50\% drop in acceptance is reached when this
misalignment reaches $\Delta r\sim 0.05r_{a}$. Lines (iii) and (iv) plot the
random and constant cases for a $(FO)^{3}(DO)^{3}$ array with
$\kappa$=38.7\,m$^{-1}$, $L$=6\,mm and $S$=24\,mm. Here, the acceptance is not
so sensitive to the random misalignments, while the sensitivity to the constant
misalignment is the same as for the $FODO$ case. These curves give some
typical scenarios. In general, we find that the transmission is less sensitive
to misalignments for smaller values of $\kappa L$, and that misalignments are
most severe when the array is operated close to the stability boundary at
$\Phi=\pi$.

In our decelerators with $r_{a}$=1\,mm, we have achieved values of
approximately 20\,$\mu$m for the size of the random misalignment by specifying
tight machining tolerances where appropriate. An alignment jig was used to
reduce the constant type of misalignment below $\sim$50\,$\mu$m.

\subsection{\label{Sec:longitudinal}
Longitudinal motion}

In order to decelerate or accelerate the molecules, time-varying
electric fields are applied. A molecule in a high-field seeking
state will gain kinetic energy as it enters the field of a lens,
while it loses kinetic energy as it leaves the lens, as shown
schematically in Fig.~\ref{Fig:ExperimentLayout}(b). If the
electric field is switched on while the molecule is inside a lens
there is no change to its kinetic energy but the molecule will
decelerate as it leaves the lens. The moment when the field is
switched on determines the effective length $L_{\mathrm\it{eff}}$
of the lens and hence the focusing properties. The moment when the
field is switched off determines the deceleration properties of
the lens. We switch off the electric fields when the molecules
have not yet left the field of a lens completely, as shown in
Fig.~\ref{Fig:ExperimentLayout}(b). This ensures that molecules at
the head of the pulse lose more kinetic energy, while those at the
tail lose less. In this way, molecules with a suitably narrow
spread of longitudinal position and velocity can be be confined to
a small area of phase-space throughout the decelerator. This
behaviour, known as phase stability, has been discussed
extensively in the context of decelerating weak-field seeking
molecules \cite{Bethlem:PRL84:5744:2000}.

\begin{figure}
\centerline{\epsffile{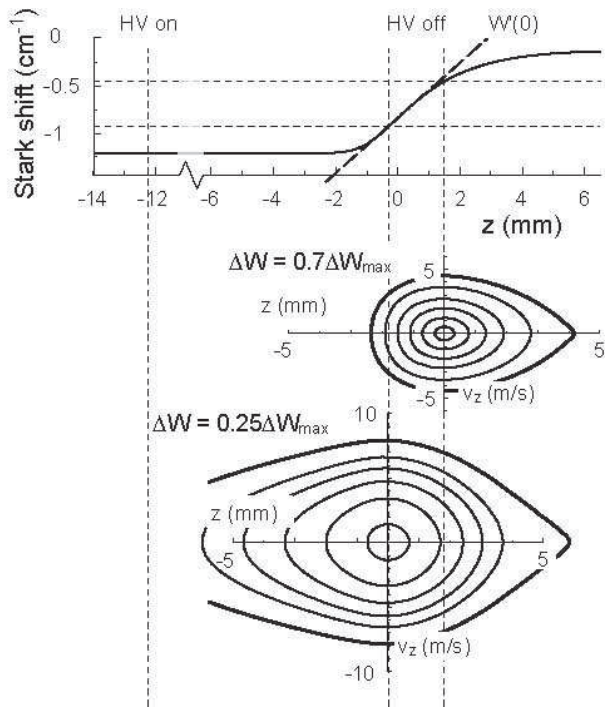}}
\protect
\vspace*{0.5cm}
\caption{\label{Fig:phasefish}
The Stark shift of metastable CO ($a^3\Pi_{1}$, $v'$=0, $J'$=1, $M\Omega$=1)
molecules as a function of their position along the molecular beam axis. The
dotted lines indicate the position of the synchronous molecule when the
electric fields are switched on and off for two different settings of the
decelerator (expressed as the energy change per stage, $\Delta W$, compared to
the maximum possible energy change, $\Delta W_{\mathrm{\it{max}}}$). In the
lower part of the figure some (closed) trajectories in phase-space are shown
for non-synchronous molecules, relative to the position and velocity of the
synchronous molecule.
}\end{figure}

To analyze the longitudinal motion, we begin by Taylor expanding the on-axis
potential energy of a single deceleration stage (see
Fig.~\ref{Fig:phasefish}) around the point of inflection at $z=0$:

\begin{equation}
W(z) = W(0) + W'(0)z + W'''(0)z^{3} + \cdots,
\end{equation}

\noindent where $W'(0)=\left.\frac{\partial W}{\partial z}
\right|_{z=0}$, $W'''(0)=\left.\frac{\partial^{3} W}{\partial
z^{3}} \right|_{z=0}$ and we have used the fact that $\partial^{2}
W/\partial z^{2}|_{z=0} = 0$. Close to $z=0$ the potential energy
can be approximated using the first two terms only. The switching
sequence is constructed such that a hypothetical molecule, the
so-called `synchronous molecule', always reaches the same position
$z_{s}$ of the relevant lens at the moment when the fields are
turned off. The change in kinetic energy of the synchronous
molecule is the same in every lens,
$W_{\mathrm{lens}}-W(z_{s})$, $W_{\mathrm{lens}}$ being the Stark shift
of the molecule inside the lens.

When the energy taken out per stage is small compared to the total
kinetic energy of the molecules \cite{Bethlem:PRL84:5744:2000},
i.e., when $\Delta v \ll v$, one can describe this change in
energy as originating from a constant force

\begin{equation}
F_{s} = \frac{W_{\mathrm{lens}}-W(z_{s})}{(L+S)}, \label{eq:averageFs}
\end{equation}

\noindent
where $(L+S)$ is the distance that the synchronous molecules travels between two
subsequent switching times. The difference of the force on a non-synchronous
molecule at position $z$ and the force on the synchronous molecule at position
$z_s$ is now given by

\begin{equation}
F - F_{s} = \frac{-W(z) + W(z_{s})}{(L+S)}
\approx -\frac{W'}{(L+S)}(z-z_{s}).
\label{eq:averageF}
\end{equation}

\noindent
Consequently, the non-synchronous molecules oscillates around the synchronous
molecule with an angular frequency given by

\begin{equation}
\omega_{z} = \sqrt{\frac{W'}{m(L+S)}},
\label{eq:axialfreq}
\end{equation}

\noindent with $m$ being the mass of the molecules.

\setlength{\epsfxsize}{0.45\textwidth}

In our experiment with metastable CO molecules, described in
Sec.~\ref{Sec:experimental}, $L+S$=30\,mm, and $W'(0)$=0.29\,cm$^{-1}$/mm
resulting in a longitudinal frequency $\omega_{z}/2\pi$=320\,Hz.

Figure~\ref{Fig:phasefish} shows the trajectories of a few
non-synchronous molecules, plotted in phase space relative to the
position and velocity of the synchronous  molecule. From these
numerical simulations, non synchronous molecules are found to
oscillate around the synchronous molecule with a frequency of
$\omega_{z}/2\pi$=330\,Hz, close to the frequency given by
Eq.~(\ref{eq:axialfreq}). The thick curves in
Fig.~\ref{Fig:phasefish} show the outermost trajectories of
molecules that are still phase stably decelerated. The
longitudinal acceptance is about 50\,mm$\cdot$m/s, when $\Delta
W=0.7\Delta W_{\mathrm{\it{max}}}$, and three times larger when
$\Delta W=0.25\Delta W_{\mathrm{\it{max}}}$.

Alternating gradient deceleration can also be applied to low-field
seeking molecules. In this case, a slightly more complicated
switching pattern must be used to achieve both longitudinal and
transverse stability. Suppose we want the synchronous molecule to
lose an energy 0.5\,$\Delta W_{\mathrm{\it{max}}}$ per stage. The
fields should be turned on well before the synchronous molecule
approaches the lens, and turned off again when it is half way up
the potential hill. This ensures that non-synchronous molecules
oscillate around the synchronous one as before. Once the molecules
are well inside the lens, the fields are turned on again to focus
the molecules, and must be turned off before they approach the
exit to ensure that they are not accelerated out of the lens. The
first high voltage pulse determines the amount of deceleration
whilst the second determines the effective length of the lens.

\subsection{\label{SubSeq:coupling}
Coupling between the longitudinal and transverse motion}

The transverse stability depends on the longitudinal velocity
because $\kappa$ is inversely proportional to $v_{z}$. Suppose the
experimental settings at the start of the decelerator correspond
to the point $(\kappa L,\kappa S)$ indicated by the dot in the
transverse acceptance plot of Fig.~\ref{Fig:acceptance}. As the
molecules are decelerated, their position on this plot moves away
from the origin along the dashed line. Eventually, this point will
move out of the region of stability, and the beam will be lost. To
avoid this, either $\Omega$ or $L$ and $S$ must be altered along
the array in sympathy with the decreasing speed. A decrease in
$\Omega$ could be achieved by decreasing the curvature of the
electric field. This could be done without altering the on-axis
field, which governs the energy loss per stage. However, the
transverse acceptance is proportional to $\Omega$, and unless the
decrease in $\Omega$ is compensated by an increase in $d$, this
will lead to beam loss. A more satisfactory approach is to
decrease $L$ and $S$ so that $L/v_{z}$ and $S/v_{z}$ remain
constant. With this approach the lenses will be long at the
beginning of the decelerator and since deceleration occurs only at
the end of each lens the overall length may then become
undesirably large. That problem can be circumvented by splitting
each lens into several parts, i.e., by replacing the $FODO$ array
with the more general $(FO)^{n}(DO)^{n}$ array. As the velocity is
decreased, $n$ is also decreased until, at the end of the
decelerator, $n=1$. In this way the beam can be decelerated stably
to a small fraction of its initial speed.

In the longitudinal direction, molecules oscillate around the
position of the synchronous molecule, causing the effective lens
length experienced by a molecule to vary according to the phase of
its longitudinal oscillation. This also couples the longitudinal
and transverse motions, possibly leading to parametric
amplification of the transverse oscillation
\cite{vandeMeerakker:PRA73:023401:2006}. The coupling can be
suppressed by designing the lenses to be long compared to the
longitudinal spread of the decelerated beam.

\section{\label{Sec:experimental}
2D Imaging of an AG decelerated beam of CO molecules}

\setlength{\epsfxsize}{0.9\textwidth}
\begin{figure*}
\centerline{\epsffile{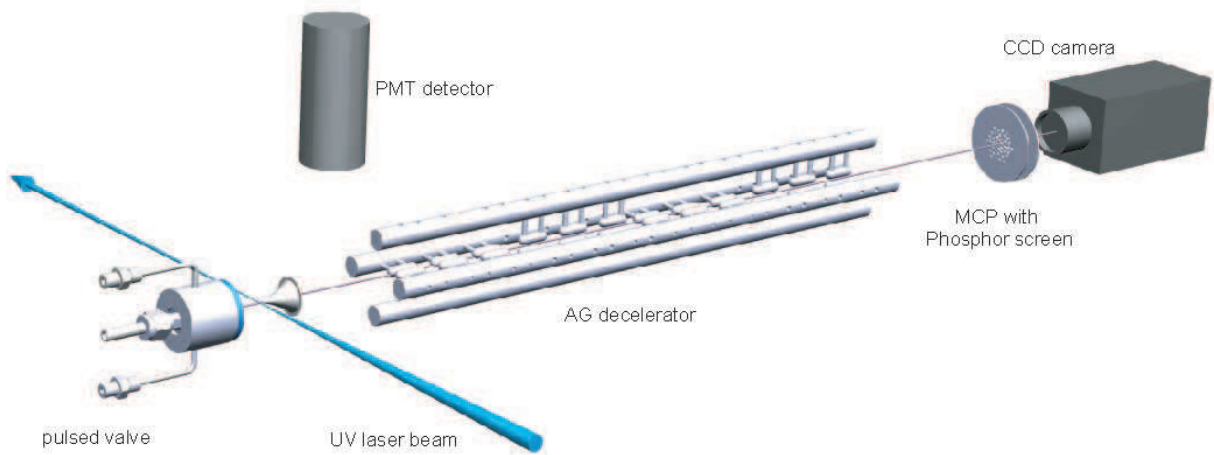}}
\protect
\vspace*{0.5cm}
\caption{\label{Fig:setup} Scheme of the experimental setup. CO
molecules are prepared by UV laser excitation to a high-field
seeking level of the metastable a$^3\Pi$ state and pass through an
array of 12 lenses arranged in the $(FO)^{3}(DO)^{3}$
configuration, with the last group of electrodes oriented
vertically. The transverse distribution of high-field seeking
metastable CO molecules is recorded 20\,cm after the decelerator
using a micro channel plate, phosphor screen and CCD camera.
}\end{figure*}

In order to demonstrate the performance of the alternating gradient
decelerator, we have carried out experiments on CO molecules in the $a^3\Pi$
state. The main reason for choosing metastable CO molecules for these
experiments is that (i) they can be prepared in a single quantum state at a
well-defined position and time, and (ii) their velocity distribution as well as
their transverse distribution can be readily recorded. A scheme of the
experimental setup is shown in Fig.~\ref{Fig:setup}. A pulsed beam of CO is
produced by expanding a mixture of CO with either Xe or Ar into vacuum, using a
modified solenoid valve. When seeded in Argon the mean velocity of the beam is
$v_{i}$=630\,m/s, corresponding to an initial CO kinetic energy of E$_{kin}$ =
480\,cm$^{-1}$. When Xenon is used and when the valve housing is cooled to
180\,K (just above the boiling point of Xe at the pressure used) the mean
velocity of the CO molecules in the beam is reduced to 315\,m/s (mixture of
20\% CO in Xe) or to 275\,m/s (mixture of 5\% CO in Xe) corresponding to
initial kinetic energies of 120\,cm$^{-1}$ and 89\,cm$^{-1}$, respectively. In
all cases, the velocity spread is approximately 10\%, corresponding to a
translational temperature of about 1\,K.

The metastable CO molecules are prepared in a single quantum state by direct
laser excitation on the spin-forbidden $a^3\Pi$ ($v'$=0) $\leftarrow$
$X^1\Sigma^+$ ($v''$=0) transition, using narrow-band pulsed 206\,nm (6.0\,eV)
radiation. In the experiments reported here, the laser is tuned to excite the
lower $\Lambda$-doublet component of the $J'$=1 $a^3\Pi_1$ level via the
$R_2(0)$ transition. By setting the polarization of the laser perpendicular to
the stray electric fields present in the excitation region only the $M\Omega=1$
high-field seeking level is prepared.

The CO molecules pass through a 1.0\,mm diameter skimmer into a second,
differentially pumped, vacuum chamber housing the 35\,cm long AG decelerator.
The decelerator consists of 12 equidistant 20\,mm long lenses, separated by
10\,mm long drift regions. The lenses are arranged in four groups of three, with
the first group of electrodes oriented horizontally and the last group of
electrodes oriented vertically. The lenses are formed from  two circular
electrodes (bold white lines of Fig.~\ref{Fig:multipole}(a)), with $r_{0}$=1\,mm
and $R$=3\,mm, and have hemispherical ends. The two opposing rods are
simultaneously switched between 0\,kV and $\pm$10\,kV by two independent high
voltage switches. The electric field on the axis is 95\,kV/cm, corresponding to
a Stark shift of -1.2\,cm$^{-1}$ for the metastable CO molecules. The Stark
shift on the molecular beam axis is shown as a function of $z$ in
Fig.~\ref{Fig:ExperimentLayout}(b).

The molecules land on a micro-channel plate (MCP) detector placed on the beam
axis. The 6~eV energy of the excited state is sufficient to release Auger
electrons from the surface. These are amplified and detected on a phosphor
screen using a CCD camera (LaVision GmbH). Thus, the 2D distribution of the
metastable CO beam is recorded. The detection efficiency of the MCP detector is
estimated to be about $10^{-3}$ \cite{Jongma:JCP102:1925:1995}. Detection
efficiencies $>10$\% can be obtained by letting the molecules impinge on a flat
gold surface kept at 500\,K and redirecting the Auger electrons towards an MCP
mounted off-axis. Unfortunately, our attempts to build sufficiently
distortion-free optics to image the electrons from the gold surface onto the
MCP detector failed. Therefore, the longitudinal characteristics of the
decelerator are recorded using the gold plate detector, but the 2D distribution
of the decelerated CO is measured with the MCP directly intercepting the beam.
The initial intensity of the metastable CO beam is monitored simultaneously by
detecting the $a^3\Pi$ $\leftarrow$ $X^1\Sigma^+$ fluorescence near the
entrance of the AG decelerator with a photomultiplier tube (PMT).

\setlength{\epsfxsize}{0.45\textwidth}
\begin{figure}
\centerline{\epsffile{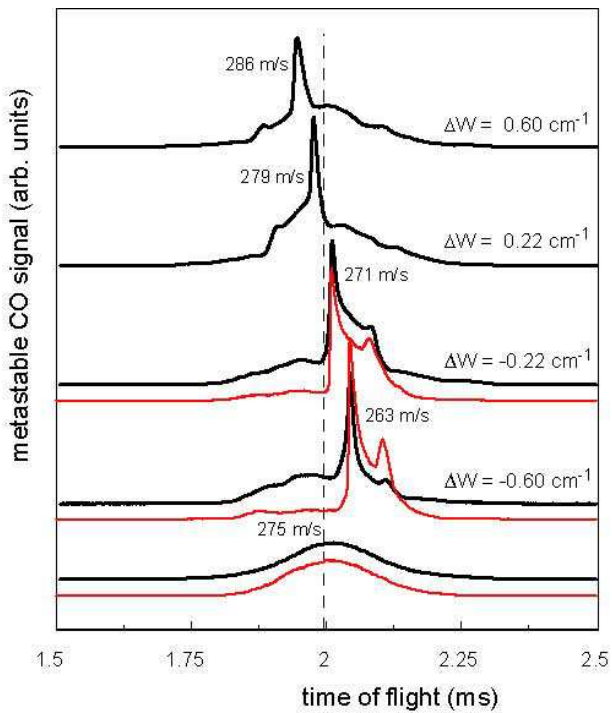}}
\protect
\vspace*{0.5cm}
\caption{\label{Fig:TOF} Observed time-of-flight (TOF) distributions
of metastable CO molecules over the 54\,cm path length through the
apparatus, for 4 different settings of the decelerator (expressed
as the energy change per stage, $\Delta W$). The lowest curve
shows the TOF-distribution when the electric fields are off. The
measurements (thick curves) have been given an offset for clarity.
The thin curves show the results of 3D trajectory calculations.
The vertical dashed line indicates the expected arrival time of a
molecule flying with a constant velocity of 275\,m/s. }
\end{figure}

Figure~\ref{Fig:TOF} shows the measured time-of-flight (TOF) distributions for
several values of the energy change per stage, $\Delta W$. The timing sequences
were chosen so that the decelerator would always act on a group of molecules
with initial speeds centred on $v_{i}=275$\,m/s. The lowest curve is the TOF
distribution obtained with no voltages applied to the decelerator. Molecules
with a speed of 275\,m/s arrive at the time indicated by the dashed line in the
figure. Using the 12 stages, the speed can be reduced to 263\,m/s, or increased
to 286\,m/s, depending on the timing of the switched fields. In these
experiments the decelerator was operated at $\pm$8\,kV. The thin curves showing
the results of the trajectory calculations discussed earlier describe the TOF
distributions for the decelerated bunch rather well. Similarly good agreement
is obtained in the simulation of the accelerated beam (not shown). The ratio of
the time-integrated signal with decelerator on and off is $\sim$1.5 as also
predicted by our simualtions. This ratio is about 10\,times larger than measured
in a previous experiment on metastable CO \cite{Bethlem:PRL88:133003:2002}.

\setlength{\epsfxsize}{0.74\textwidth}
\begin{figure*}
\centerline{\epsffile{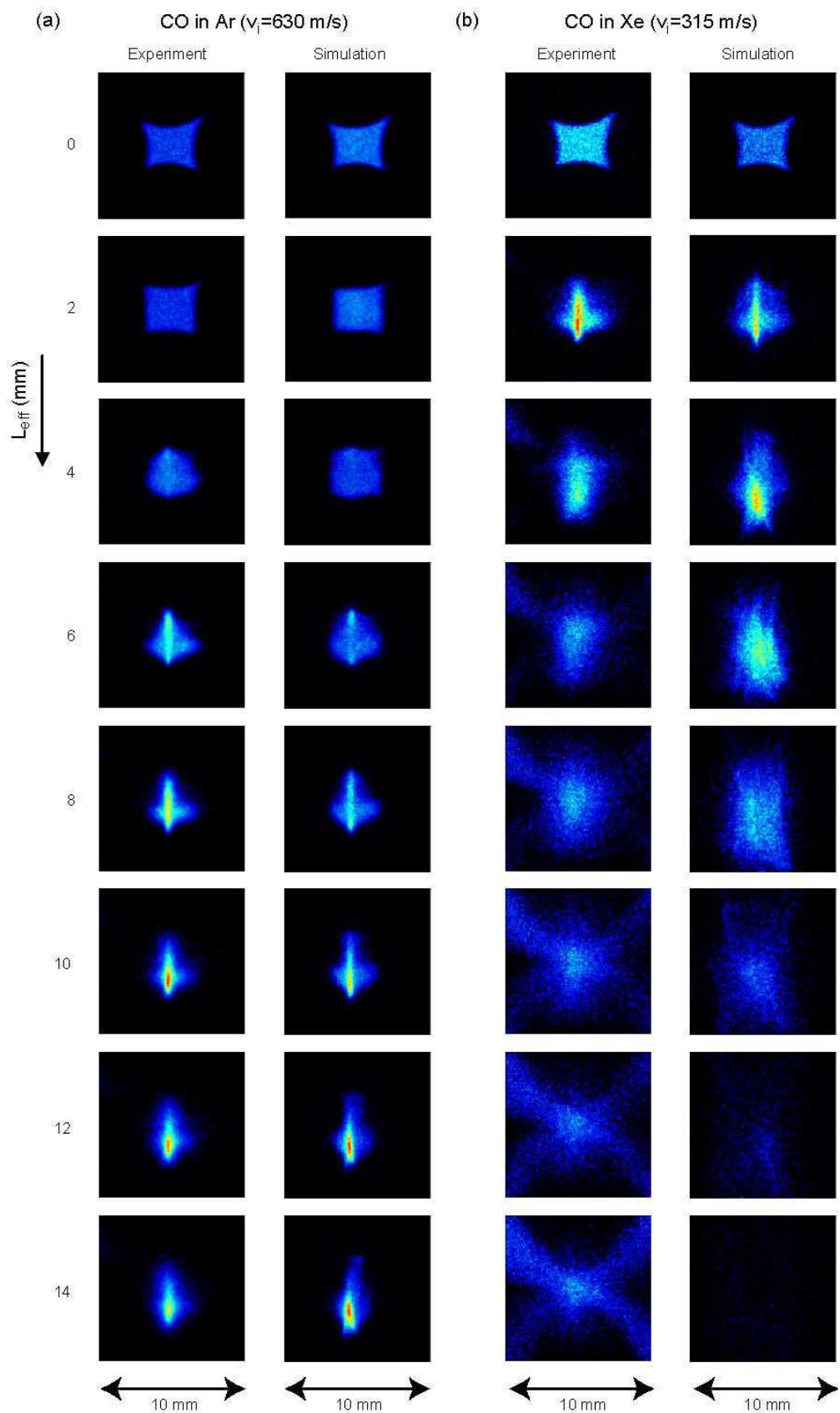}}
\protect
\vspace*{0.5cm}
\caption{\label{Fig:2D} Observed and calculated 2D distributions
of a decelerated beam of CO molecules with an initial speed of (a)
630\,m/s and (b) 315\,m/s, for various values of the effective lens
length (in mm). }\end{figure*}

Figure~\ref{Fig:2D} shows transverse distributions of the
molecules, measured 20\,cm downstream from the decelerator exit
for a variety of effective lens lengths. The imaging detector is
switched on for a short period (20\,$\mu$s when $v_{i}$=630\,m/s
and 40\,$\mu$s when $v_{i}$=315\,m/s) so that only the decelerated
molecules are detected. The images are formed from $\sim 10^{4}$
detected molecules accumulated over many shots (2000 when
$v_{i}$=630\,m/s and 10$^{4}$ when $v_{i}$=315\,m/s). The false
colour is a measure of the number of molecules detected in each
pixel. A calculated spatial distribution is shown beside each
experimental image.

We discuss first the data with $v_{i}=630$\,m/s, Fig.~\ref{Fig:2D}(a). The top
image shows the 2D distribution obtained when there are no voltages on the
decelerator. As expected, this measurement simply shows equally distributed
molecules within the aperture formed by the electrodes. The detected
distribution has a width-to-height ratio of about 1.3 because the last 3
electrode pairs are oriented vertically and end 9\,cm closer to the detector
than the last horizontal electrode pairs. By looking at the outline of the
electrodes, one can also see that the lens array was not perfectly aligned.
Indeed, the centre of the horizontal lens set is displaced by 250\,$\mu$m from
the center of the vertical set (we have since reduced this misalignment to less
than 50\,$\mu m$). This misalignment has been included in the simulations.

The remaining images in Fig.~\ref{Fig:2D}(a) show how the profile
of the beam changes as the effective length of the lenses is
increased. We consider the vertical direction first. In this
direction, the last set of lenses are defocussing, and the beam is
diverging when it exits the decelerator. One might expect this
divergence to increase with increasing lens length. However, the
smaller size of the beam inside the defocussing lenses (compare
(a) and (b) in Fig.~\ref{Fig:trajectories}) tends to compensate
for the increased power of those lenses. As a result, the vertical
divergence changes very little in our experiments and the height
of the distribution is approximately constant in all the images.
In the horizontal direction, the last set of lenses are focusing
lenses, and so the beam is converging when it exits the
decelerator. In contrast to the defocussing lenses, the size of
the beam inside the focussing lenses is fixed, being determined by
the lens aperture. As the lens length increases the exiting beam
converges more strongly and so the width of the detected
distribution decreases. When the lens length is about 8\,mm, a
focus is formed in the plane of the detector. Here, the settings
correspond to the dot placed on the transverse acceptance plot of
Fig.~\ref{Fig:acceptance}(b), where the phase-advance is $\pi/6$.
The focus that is formed is rather aberrant, resembling a cross
rather than a vertical line. As we will see shortly, the
aberration is caused by non-linearities in the transverse forces.
When the length is increased beyond 8\,mm the focus lies in
frontupstream of the detector and the horizontal width begins to
increase again. In all cases, the experimental images agree very
well indeed with the simulations.

Turning now to the data with $v_{i}$=315\,m/s (Fig.~\ref{Fig:2D}(b)), we see
that the lower forward speed results in the focus being formed in the plane of
the detector for a smaller value of $L_\mathrm{\it{eff}}$. Using a thin lens
approximation, the focal length is $1/\kappa^{2}L$. Since halving $v_{i}$,
doubles $\kappa$ we expect the focus to be formed for
$L_\mathrm{\it{eff}}$=2\,mm, exactly as observed in the data. These settings
correspond to the cross placed on Fig.~\ref{Fig:acceptance}(b), where again, the
phase advance is $\pi/6$. As $L_\mathrm{\it{eff}}$ is increased beyond 2\,mm,
the horizontal focus lies progressively further in front of the detector and so
the detected distribution grows progressively wider. When
$L_\mathrm{\it{eff}}>10$\,mm, the trajectories become unstable and there is a
very sudden drop in the intensity of the simulated data. The experimental
images show a less rapid drop in intensity, CO molecules being observed near
the molecular beam axis, even when $L_\mathrm{\it{eff}}=14$\,mm. In addition, a
diagonal cross shape is observed in the experimental data. This we attribute to
molecules so strongly focused in the first three lenses that they escape and
subsequently fly outside the decelerator. Our simulation could not follow such
trajectories because the field used was bounded 4\,mm from the beam axis and
molecules outside this area were considered lost. We surmise that these
molecules also give rise to the intensity observed at the centre of the images.

By integrating the intensity of the measured distributions one obtains the
acceptance as a function of the effective length of the lenses, analogous to
the calculations shown in Fig.~\ref{Fig:online}. However, a frequency drift of
the UV excitation laser during the measurements compounded by the rather low
counting rates, resulted in integrated beam fluctuations of more than 50\%,
preventing a clear comparison of measured and calculated acceptances.

\setlength{\epsfxsize}{0.45\textwidth}
\begin{figure}
\centerline{\epsffile{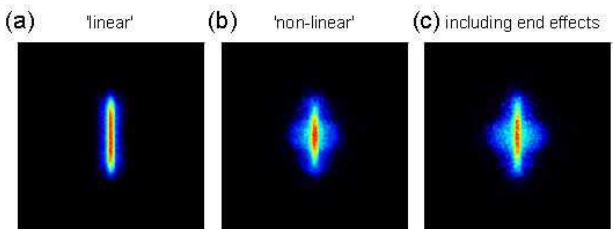}}
\protect
\vspace*{0.5cm}
\caption{\label{Fig:nlin2D} Simulated 2D distributions 20\,cm
downstream from the decelerator exit, for CO molecules with
$v_{i}=630$\,m/s. (a) Linear force, (b) true force, excluding end
effects, and (c) true force.}\end{figure}

Figure~\ref{Fig:nlin2D} reveals the origin of the aberrations present at the
focus. We repeated the calculation for the case of $v_{i}=630$\,m/s,
$L_\mathrm{\it{eff}}=8$\,mm. In (a), the calculation was performed assuming a
perfectly linear force, and we find the focus to be a vertical line without
aberration. The aberrations appear when we include the non-linearities in the
force (Fig.~\ref{Fig:nlin2D}(b)), showing this to be the primary factor in
degrading the image quality. The addition of the end-effects degrades the image
quality a little further, as shown in (c).

\section{\label{Sec:conclusion}
Summary and conclusions}

In this paper we have presented the principles of alternating gradient
deceleration of polar molecules along with the criteria that govern decelerator
design and operation. We began by showing that heavy molecules in low-lying
rotational states seek strong field, and so cannot be confined by static
electric fields. Alternating gradient focusing can be used to stabilize the
trajectories of such molecules in a Stark decelerator. We showed how the
electric fields required to achieve alternating gradient focusing can be
produced using simple electrode geometries. The simplest of these, the two-rod
geometry shown in Fig.~\ref{Fig:multipole}(a), has already been used to
decelerate metastable CO and YbF molecules
\cite{Bethlem:PRL88:133003:2002,Tarbutt:PRL92:173002:2004}, and was used in the
imaging experiments presented here. This electrode geometry gives rise to
significant non-linear terms in the force. These can be reduced, without
reducing the on-axis field, by the addition of a pair of grounded electrodes.
The non-linearities are further reduced when four high voltage electrodes are
arranged as in Fig.~\ref{Fig:multipole}(b), but at the cost of reducing the
field on the axis. The fringe-fields of the lenses, essential for deceleration,
tend to increase the defocusing power relative to the focusing power near the
entrance and exit of the lens. This detrimental effect can be reduced by
terminating the rods more gradually. The four-rod geometries are also more
favourable than the two-rod geometry in this respect.

We discussed the motion of the molecules through the decelerator in detail.
When the transverse forces are perfectly linear, the transverse motion is best
described using the formalism first set out in the context of the alternating
gradient synchrotron \cite{Courant:AnnPhys3:1:1958}, and outlined in
Sec.~\ref{subSec:transverse}. The trajectories are described in terms of the
phase-advance and the envelope function. The transverse phase-space
distribution is an ellipse whose shape evolves periodically through the
lattice, but whose area is a constant. Figure~\ref{Fig:trajectories}
illustrates the evolution of the beam envelope and the phase-space ellipse. We
calculated the transverse acceptance of an array of ideal lenses as a function
of $\kappa L$ and $\kappa S$ (Fig.~\ref{Fig:acceptance}) and found the maximum
acceptance to be $0.186d^{2}\Omega$, where $d$ is the transverse aperture and
$\Omega$ given in terms of the mass, the effective dipole moment and the
electric field curvature by Eq.~(\ref{Eq:omega}). Non-linearities in the
transverse forces, and the fringe-fields of the lenses, severely reduce the
transverse acceptance (Fig.~\ref{Fig:online}). Furthermore the degree of
lens-to-lens alignment required to achieve a high transmission was found to be
demanding but attainable with high-precision machining.

The longitudinal motion was discussed in terms of a simple model of
phase-stability. The most significant coupling between transverse and
longitudinal motion is that the transverse motion depends on the forward speed.
As the speed is reduced the molecules spend more time in each lens. A good way
to handle this is to use an $(FO)^{n}(DO)^{n}$ structure with $n$ larger at
the entrance than at the exit of the decelerator.

We studied the focusing properties of an alternating gradient decelerator
experimentally using an array of 12 lenses, by measuring 2D images of a
decelerated beam of metastable CO molecules. Trajectory simulations of this
experiment reproduce the experimental findings. Non-linearities in the force
and effects due to rounded ends of the electrodes need to be included to obtain
close agreement. These experimentally verified simulations predict a transverse
acceptance of 2\,(mm.m/s)$^2$, see Fig.~\ref{Fig:online}, and a longitudinal
acceptance of 50\,mm.m/s, see Fig.~\ref{Fig:phasefish}. These numbers can be
compared to those that have been presented earlier for light molecules in
low-field seeking states. In the Stark deceleration of ammonia, for instance, a
transverse acceptance of 160\,(mm.m/s)$^2$ and a longitudinal acceptance of
10\,mm.m/s has been obtained \cite{Bethlem:PRA65:053416:2002}. As discussed
above, the transverse acceptance is very much smaller than for ideal lenses and
we can expect a more sophisticated lens design to yield a tenfold improvement.
A further increase of the acceptance could be achieved by increasing the
transverse aperture, although a corresponding increase in the applied voltages
would be needed in order to maintain the same on-axis electric field.

In discussing the alternating gradient decelerator, we have also laid out the
principles of a guide for high-field-seeking molecules. Unlike a decelerator, a
guide does not need to be divided into segments along the beamline, and so can
be free of end-effects. The symmetric four electrode geometry of
Fig.~\ref{Fig:multipole}(b) makes an ideal guide because its aberrations are
small, its acceptance is high, and the focus and defocus directions are very
easily switched. Another interesting application is to use this geometry as an
$m/\mu_{\mathrm{\it{eff}}}$ filter, the equivalent of an $m/q$ filter for ions
\cite{Paul:RMP62:531:1990}. The resolution of such a filter can be increased,
at the expense of the acceptance, by tuning the focus-defocus duty-cycle away
from 50\%. In one plane, the defocusing lenses are then longer than the
focusing lenses making the stability region narrower and so increasing the
resolution. Calculations indicate that a resolution
$\Delta(m/\mu_{\mathrm{\it{eff}}})/(m/\mu_{\mathrm{\it{eff}}}) \sim 0.1$ can be
obtained at the cost of a factor of 4 in acceptance relative to the maximum.

The decrease in velocity achieved in an AG decelerator so far has been rather
small, but since the trajectories of the molecules through the decelerator are
inherently stable, no additional losses are expected when the number of stages
is further increased. For the molecules listed in Table~\ref{Tab:Stark}, and
for many others, approximately 100 electric field stages are sufficient to
bring them to rest. These molecules could subsequently be stored in a storage
ring \cite{Nishimura:EJPD31:359:2004} or in an AC-trap
\cite{vanVeldhoven:PRL94:083001:2005}.

\section{Acknowledgements}

We acknowledge the expert technical assistance of H. Haak and J. Dyne. We thank
B. Friedrich and J. van Veldhoven for helpful discussions and S.Y.T. van de
Meerakker and F. Filsinger for help with performing the calculations of the
Stark effect. This work was supported by the `Cold Molecules' network of the
European Commission and, in the UK, by EPSRC and PPARC. H.L.B acknowledges
financial support from the Netherlands Organisation for Scientific Research
(NWO) via a  VENI-grant. M.R.T and E.A.H acknowledge the support of the Royal
Society.

\end{document}

%% file: Bethlem_Starktable.tex
\begin{footnotesize}
\begin{tabular}{llcccc} \hline\hline
\\
Molecule
& Rotational state
& Stark shift
& Effective dipole ~
&
Rotational constants ~
& Mass \\
&
& (cm$^{-1}$)
& (cm$^{-1}$/kV/cm)
& (cm$^{-1}$)
& (amu) \\
&
& at 100 kV/cm ~
& at 100 kV/cm ~
& A~/~B~/~C
&
\\
\hline
&~&~&~&~&\\

CO $\left(\emph{a}\,^{3}\Pi_{1} \right)$ \cite{Jongma:CPL270:304:1997}
& $ |J=1, M\Omega=-1\rangle$
& $-$1.25
& 0.0135
&  -~/~1.68~/~-
&  28
\\[2ex]

CaF \cite{Childs:JCP80:2283:1984, Kaledin:JMolSpec197:289:1999}
& $ |J=1/2, M\Omega=+1/4 \rangle$
& $-$3.43
& 0.0420
&  -~/~0.34~/~-
& 59
\\[2ex]

YbF \cite{Sauer:JCP105:7412:1996}
& $ |J=1/2, M\Omega=+1/4 \rangle$
& $-$4.91
& 0.0569
& -~/~0.24~/~-
& 193
\\[2ex]

ND$_{3}$ \cite{Baugh:CPL:219:207:1994}
& $|J=1,MK = -1 \rangle$
& $-$1.27
& 0.0134
&  -~/~5.14~/~3.12
& 20
\\[2ex]

pyridazine \cite{Li:JPCA102:8084:1998}
& $|J_{K_{a}K_{c}} \vert M \vert \rangle=|0_{00}0\rangle$
& $-$5.59
& 0.0624
& 0.21~/~0.20~/~0.10
& 80
\\[2ex]

benzonitrile \cite{Borst:CPL350:485:2001}
& $|J_{K_{a}K_{c}} \vert M \vert \rangle=|0_{00}0\rangle$
& $-$6.71
& 0.0711
& 0.19~/~0.051~/~0.040
& 103
\\[2ex]

tryptophan \cite{Compagnon_JACS123:8440:2001} ~I
& $|J_{K_{a}K_{c}} \vert M \vert \rangle=|0_{00}0\rangle$
&  $-$6.25
&  0.0646
&  0.041~/~0.013~/~0.012
&  216
\\
\hspace*{7em} II
&
&  $-$4.72
&  0.0494
&  0.039~/~0.014~/~0.012
&
\\
\hspace*{7em} III
&
&  $-$1.71
&  0.0183
&  0.033~/~0.017~/~0.013
&
\\
\hspace*{7em} IV
&
&  $-$11.68
&  0.120
&  0.032~/~0.016~/~0.013
&
\\
\hspace*{7em} V
&
&  $-$12.28
&  0.126
&  0.043~/~0.011~/~0.0096
&
\\
\hspace*{7em} VI
&
&  $-$11.37
&  0.116
&  0.045~/~0.011~/~0.0095
&
\\[2ex]

\hline\hline
\end{tabular}
\end{footnotesize}